\documentclass[twocolumn,amssymb,amsmath,
nofootinbib,tightenlines,showpacs,floatfix,
superscriptaddress,aps,prb]{revtex4-1}
\usepackage{amsfonts}
\usepackage{txfonts} 
\usepackage{amsbsy}
\usepackage{bm}%
\usepackage[colorlinks=true,linkcolor=blue,filecolor=blue,menucolor=yellow,urlcolor=blue,citecolor=blue,anchorcolor=blue]{hyperref}
\usepackage{graphicx}
\usepackage{times} 
\usepackage{dcolumn}

\begin{document}

\title{Tunneling Conductance and Spin Transport in Clean Ferromagnet-Ferromagnet-Superconductor 
Heterostructures} 

\author{Chien-Te Wu}
\email{wu@physics.umn.edu}
\author{Oriol T. Valls}
\email{otvalls@umn.edu}
\altaffiliation{Also at Minnesota Supercomputer Institute, University of Minnesota,
Minneapolis, Minnesota 55455}
\affiliation{School of Physics and Astronomy, University of Minnesota, 
Minneapolis, Minnesota 55455}
\author{Klaus Halterman}
\email{klaus.halterman@navy.mil}
\affiliation{Michelson Lab, Physics
Division, Naval Air Warfare Center, China Lake, California 93555}

\date{\today}

\begin{abstract} 
We present a  transfer matrix approach  
that combines
 the
Blonder-Tinkham-Klapwijk (BTK)  formalism and self-consistent
solutions to the Bogolibuov-de Gennes (BdG) equations 
and  use it  
to  study 
the 
tunneling conductance and spin transport in ferromagnet 
(${\rm F}$)-superconductor (${\rm S}$) 
trilayers (${\rm F_1F_2 S}$) as  functions of bias voltage.
The self-consistency  ensures that  
the
spin and charge conservation laws are properly satisfied.
We consider forward and angularly averaged conductances over a broad range of
the strength of the exchange fields and ${\rm F}$ thicknesses, 
as the relative in-plane magnetization angle, $\phi$, between the two 
ferromagnets varies. 
The $\phi$-dependence of the self-consistent conductance 
curves in the trilayers can differ substantially from that obtained 
via a non-self-consistent  approach. 
The zero bias forward conductance peak  
exhibits, as $\phi$ varies, resonance 
effects  intricately associated with particular 
combinations of the geometrical and material parameters. 
We find,  when the magnetizations are non-collinear, 
signatures of the anomalous Andreev reflections in the subgap 
regions of the angularly averaged conductances. 
When ${\rm F_1}$ is half-metallic, 
the angularly averaged subgap conductance 
chiefly arises from  anomalous Andreev reflection. 
The in-plane components of the spin current are strongly bias dependent,
while
the out-of-plane spin current component is only weakly 
dependent upon voltage.
The components of the spin current aligned with
the local 
exchange field of one of the F layers are 
conserved in 
that layer and in the S region, while they oscillate in the other layer.
We compute the spin transfer torques, in connection with the oscillatory
behavior of spin currents, and verify that the spin 
continuity equation is strictly obeyed in our method.
\end{abstract}

\pacs{74.45.+c,74.78.Fk,75.75.-c}  

\maketitle

\section{Introduction} 
\label{intro} 
Over the last two decades, significant progress 
in fabrication techniques has allowed
the development of  spintronics devices, such as spin 
valves,\cite{igor} that
utilize both charge and spin degrees of freedom.
Traditional spin valves  consist of magnetic 
materials only. There is another   
important type of spintronics devices, involving
ferromagnet (F)-superconductor (S) 
heterostructures.
These heterostructures have also received much attention because
of the fundamental
physics related to the 
interplay between ferromagnetic and superconducting
order. 
Their potential applications in spintronics 
include magnetic memory technology where information storage 
is accomplished via  control of 
the magnetic moment bit. It is then crucial to have 
precise
control over the magnetization direction. 
Spin transfer
torque  (STT) is one effect that
affords such control.
The generation of spin-polarized supercurrents may be used
to obtain a superconducting STT acting on the
magnetization of a ferromagnet.
This effect may  be
utilized in high density nanotechnologies that require 
magnetic tunnel junctions.
Thus, the dissipationless nature of
the supercurrent flow offers a promising avenue in terms of
low energy nanoscale manipulation of superconducting and magnetic orderings.

Although ferromagnetism and $s$-wave superconductivity 
seem incompatible 
because of the inherently opposite natures of their 
order parameter spin configurations,  superconductivity
can still be induced in the F layers of F-S layered structures by the
superconducting proximity effects.\cite{Buzdin2005} 
In essence, 
the superconducting proximity effects 
describe the leakage of 
superconductivity into a non-superconducting
normal 
(N) 
or  magnetic metal,
as well
as its depletion in S near the
interface. However,  
proximity effects in F-S systems are very different from  those in 
N-S structures due
to the inherent exchange field in the F materials.
As a consequence of this exchange field, the Cooper pair acquires a
non-zero center-of-mass momentum\cite{demler,Buzdin2005,Halterman2001,Halterman2002}
and the overall Cooper
pair wavefunction oscillates spatially in the F regions.
Owing to this oscillatory nature, many new physical phenomena emerge  
in  F-S heterostructures such as oscillations of the superconducting
transition temperature, $T_c$, with the thickness of the F layers. 
\cite{Buzdin1990,Buzdin2005,demler,Halterman2004}
 
It is of fundamental importance that superconducting proximity effects 
are governed by Andreev reflection,\cite{Andreev}
which is a process of  electron-to-hole  
conversion at N-S or F-S interfaces, and it involves the creation or 
annihilation of a Cooper pair. 
Therefore, consideration of Andreev reflection is central when studying
the transport properties of N-S\cite{btk,tanaka} 
or F-S systems.\cite{beenakker,zv1,zv2}
Of particular interest\cite{btk,tanaka,beenakker,zv1,zv2} is
the behavior of the tunneling conductance in the subgap region, 
where hybrid systems can carry a supercurrent due to Andreev reflection.
In  conventional Andreev reflection, the  reflected hole has 
opposite spin to the incident particle. Accordingly, the exchange 
field in the F materials that causes the splitting of spin bands has a 
significant effect on the tunneling conductance in the subgap region.
Most important, the qualitative behavior of the conductance peak in 
the zero bias limit is strongly influenced by the degree of conduction
electron spin polarization in the F materials.\cite{beenakker,zv1,zv2,mazin}
Experimentally, this concept has been applied to quantify 
the spin polarization.
\cite{raychaudhuri,upad,chalsani,soulen,hacohen}

An intriguing phenomenon in F-S structures is the induction of 
triplet pairing 
correlations.\cite{berg86,Bergeret2007,Wang2010,Hubler2012,hv2p}
These correlations are very important when 
studying transport phenomena such as those found in SFS Josephson 
junctions.\cite{Keizer2006,robinson,khaire}
In contrast to the short proximity 
length\cite{Halterman2002} of singlet Cooper pair condensates into F materials,
the $m=\pm1$ triplet pairing correlations are compatible
with 
the exchange fields and hence largely immune to 
the pair
breaking effect produced by the latter. 
However, for such correlations to be induced 
F-S structures must possess a spin-flip mechanism.
Examples include a spin-dependent scattering potential at the F-S interface
\cite{Halterman2009,Eschrig2008}
and the introduction of another magnetic layer with a misoriented
magnetic moment such as
${\rm F_1SF_2}$ superconducting spin valves.\cite{Halterman2007}
The  
pairing state of $m=\pm1$ induced triplet correlations 
is at 
variance with the effects of {\it conventional}
Andreev reflection, responsible for the generation 
of  singlet
Cooper pairs. Thus, recent studies\cite{linder2009,visani,niu,ji,feng} 
on the tunneling conductance propose the existence of  {\it anomalous} 
Andreev reflection, that is, a reflected
hole with the same spin as the incident 
particle can be Andreev reflected under 
the same circumstances as the generation of 
$m=\pm1$ triplet pairing correlations becomes possible. In this
view,  triplet 
proximity effects are correlated with the process of this 
anomalous Andreev reflection. This
will be confirmed and discussed in this work. 

Another important geometry for a superconducting spin valve consists
of a 
conventional spin valve with a superconductor layer
on top: a 
${\rm F_1F_2S}$ trilayer. By applying an external magnetic field, 
or switching via STT,
one is able to control the 
relative orientation of the intrinsic magnetic moments
and investigate the dependence\cite{alejandro,leksin,zdravkov} 
of physical properties
such as 
$T_c$ on the misorientation angle $\phi$ between the two magnetic layers.
Due to the proximity effects, $T_c$ is often found to be minimized when the 
magnetizations are approximately perpendicular to each other,\cite{cko} 
reflecting the presence of long range triplet correlations,
induced in ${\rm F_1F_2S}$ trilayers. 
Their existence has
been verified 
both theoretically\cite{cko} and experimentally.\cite{alejandro}
The non-monotonic behavior of $T_c$ as 
a function of $\phi$ 
has also been shown to be quantitatively\cite{alejandro}  related to  the  
long range triplet correlations, 
with excellent agreement  between
theory and experiment. 

Motivated by these important findings, 
we will investigate here, in a fully self-consistent manner, the $\phi$ 
dependence of  
the tunneling conductance and other transport quantities of these 
${\rm F_1F_2S}$ trilayers. 
Non-self-consistent theoretical studies of tunneling conductance have 
been performed on 
${\rm F_1F_2S}$ trilayers in previous work.\cite{ji,cheng} However,
as we shall see in Sec.~\ref{methods}, only
self-consistent methods guarantee that conservation
laws are not violated and (see Sec.~\ref{results}) only then 
can one correctly predict the 
proximity effects on the angular dependence
of transport properties.
The spin-polarized tunneling conductance of F-S bilayers only, 
was studied in Refs.~\onlinecite{zv1,zv2,kashiwaya,yamashita}.
Also, in  traditional spin valves 
e.g.
${\rm F_1}$-${\rm F_2}$ layered structures, the spin-polarized current 
generated in the ${\rm F_1}$ layer can transfer angular momentum to 
the ${\rm F_2}$ layer
when their magnetic moments are not parallel to each other\cite{igor} 
via the effect of STT.\cite{berger,myers} 
As a result, the spin current is not a conserved 
quantity and one needs a general law that relates 
local spin current to local STT.\cite{linder2009} 
The transport properties 
of ${\rm F_1SF_2}$ structures, in particular the
dependence on applied bias of the spin-transfer torque and
the spin-polarized tunneling conductance 
have been previously studied.\cite{romeo,bozovic,linder2009}

Here, we consider charge transport and both  spin current 
and spin-transfer torque in  ${\rm F_1F_2S}$ trilayers.
In  previous theoretical work, such as that 
mentioned above, when computing 
tunneling conductance of N-S and F-S structures, using methods based on the
Blonder-Tinkham-Klapwijk (BTK) 
procedure\cite{btk,tanaka,zv1,zv2,bozovic,linder2007,linder2009}
and quasi-classical approximations,\cite{Eschrig2010}
the superconducting pair amplitude 
was assumed to be a step function: a constant
in S,  dropping abruptly to zero at the N-S or F-S
interface and then vanishing in the non-superconducting region. 
This assumption neglects  proximity effects.  
Only  qualitative predictions on the behavior of the tunneling conductance 
can be reliably made. Still, results exhibit
many interesting features especially in F-S systems.\cite{zv1,zv2}
However, to fully account for the proximity effects,
in the  transport properties, 
one must use a self-consistent pair potential. This is because
that reveals realistic 
information regarding the leakage and depletion of superconductivity.
Also, as we shall discuss below, self-consistent solutions 
guarantee that conservation laws are satisfied. 
In Ref.~\onlinecite{bv}, the  
tunneling conductance of F-S bilayers was extracted
via 
self-consistent solutions of Bogoliubov-de Gennes (BdG) equations.\cite{degennes}
However, the numerical methods used there required awkward fitting 
procedures that led to appreciable uncertainties and precluded their 
application to trilayers. 
The findings indicated that the self-consistent tunneling conductances 
for the bilayer 
are quantitatively different from those computed in a non-self-consistent 
framework, thus
demonstrating the importance of properly accounting for proximity effects 
in that situation.  
Here we report on a powerful  {\bf self-consistent} approach and use it 
to compute the tunneling conductance of 
${\rm F_1F_2S}$ trilayers. It is 
based on the BTK method, incorporated into a transfer matrix procedure
similar to that used\cite{strinati} in Josephson junction
calculations and  simple F-S junctions within a Hubbard model\cite{ting}. 
As we shall demonstrate, 
this approach not only has the advantage of being
more numerically efficient but also can be used 
to compute spin 
transport quantities. 
Thus, we are able to address 
many important points regarding 
both charge and spin transport in ${\rm F_1F_2S}$ trilayers,
including the spin currents and spin-transfer torque,
the proximity effects on the tunneling conductance, and the correlation
between the anomalous Andreev reflection and the triplet correlations.

This paper is organized as follows: 
we  present our self-consistent approach,
and its application  to compute the tunneling conductance,  the 
spin-transfer torques,  the spin current, and the
proper way to ensure that conservation laws
are satisfied, in Sec.~\ref{methods}. 
In Sec.~\ref{results} we
present the results.
In Subsec.~\ref{bilayers}, we briefly  compare the results of F-S bilayers 
obtained in our self-consistent approach with non-self-consistent ones.
The rest of Subsec.~\ref{trilayers} includes our results for trilayers,
that is, the main results of this work. The 
dependence on the tunneling conductance of 
${\rm F_1F_2S}$ trilayers on the angle $\phi$ 
is extensively discussed
as a function of geometrical and material parameters. 
Results for the effect of the anomalous Andreev reflection, 
the spin-transfer torque, 
and the spin current are also presented. 
We conclude with
a recapitulation  in Sec.~\ref{conclusions}. 

\begin{figure}
\includegraphics[width=0.45\textwidth] {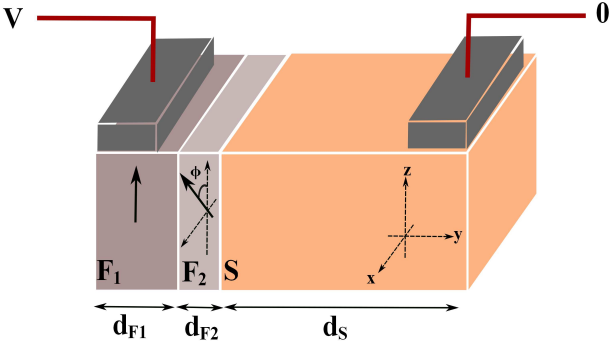} 
\caption{(Color online) Schematic of the F$_1$F$_2$S trilayer. The
exchange field, $\bm h$, denoted by a black solid arrow, 
is along the $+z$ direction in the outer magnetic layer (F$_1$)
while within the inner magnetic layer 
(F$_2$), $\bm h$ is oriented at an angle $\phi$ in the $x-z$ plane. 
The outer magnetic layer and the superconducting layer
are connected to electrodes that are biased with a finite voltage $V$.}
\label{figure1}
\end{figure}

\section{Methods}
\label{methods}
\subsection{Description of the system}
\label{description}
The geometry of our system  is depicted in Fig.~\ref{figure1}.
We denote the outer ferromagnet as ${\rm F_1}$ and the middle layer as
${\rm F_2}$.
We choose our coordinate
system so that  the interfaces are parallel to the $x-z$ plane, and
infinite in extent, while  the system
has a finite width $d=d_{F_1}+d_{F_2}+d_S$ in the $y$ direction. 

The Hamiltonian appropriate to our system is,
\begin{eqnarray}
\label{ham}
{\cal H}_{eff}&=&\int d^3r \left\{ \sum_{\alpha}
\psi_{\alpha}^{\dagger}\left(\mathbf{r}\right){\cal H}_0
\psi_{\alpha}\left(\mathbf{r}\right)\right.\nonumber \\
&+&\left.\frac{1}{2}\left[\sum_{\alpha,\:\beta}\left(i\sigma_y\right)_{\alpha\beta}
\Delta\left(\mathbf{r}\right)\psi_{\alpha}^{\dagger}
\left(\mathbf{r}\right)\psi_{\beta}^{\dagger}
\left(\mathbf{r}\right)+H.c.\right]\right.\nonumber \\
&-&\left.\sum_{\alpha,\:\beta}\psi_{\alpha}^{\dagger}
\left(\mathbf{r}\right)\left(\mathbf{h}\cdot\bm{\sigma}
\right)_{\alpha\beta}\psi_{\beta}\left(\mathbf{r}\right)\right\},
\end{eqnarray}
where ${\cal H}_0$ is the single-particle Hamiltonian, $\mathbf{h}$
is a Stoner exchange field that characterizes the magnetism,
and $\bm{\sigma}$
are Pauli matrices.
The superconducting pair potential $\Delta(\mathbf{r})\equiv 
g\left(\mathbf{r}\right)\left\langle\psi_{\uparrow}
\left(\mathbf{r}\right)\psi_{\downarrow}\left(\mathbf{r}\right)  
\right\rangle$ is the product of pairing constant,
$g\left(\mathbf{r}\right)$, in the singlet channel, and
the pair amplitude.
We begin by writing down the BdG equations,
which we will solve self-consistently for our F$_1$F$_2$S trilayers. 
By performing the generalized Bogoliubov transformation\cite{degennes}, 
$\psi_{\sigma}=\sum_n\left(u_{n\sigma}\gamma_n 
+\eta_{\sigma}v_{n\sigma}^{\ast}\gamma_n^{\dagger}\right)$,
where $\sigma= (\uparrow, \downarrow)$  and $\eta_{\sigma}\equiv1(-1)$ for spin-down (up),
the Hamiltonian [Eq.~(\ref{ham})] can be diagonalized.  
We can then for our geometry
rewrite\cite{cko}
Eq.~(\ref{ham}) as a quasi-one-dimensional eigensystem: 
\begin{align}
&\begin{pmatrix}
{\cal H}_0 -h_z&-h_x&0&\Delta \\
-h_x&{\cal H}_0 +h_z&\Delta&0 \\
0&\Delta&-({\cal H}_0 -h_z)&-h_x \\
\Delta&0&-h_x&-({\cal H}_0+h_z) \\
\end{pmatrix}
\begin{pmatrix}
u_{n\uparrow}\\u_{n\downarrow}\\v_{n\uparrow}\\v_{n\downarrow}
\end{pmatrix} \nonumber \\
&=\epsilon_n
\begin{pmatrix}
u_{n\uparrow}\\u_{n\downarrow}\\v_{n\uparrow}\\v_{n\downarrow}
\end{pmatrix}\label{bogo},
\end{align}
where the $u_{n\sigma}$ and $v_{n\sigma}$ are respectively the quasiparticle 
and quasihole amplitudes with spin $\sigma$. 
The  exchange
field vanishes in the S region, while 
in ${\rm F_1}$ 
it is directed  along $z$, $\mathbf{h}=h_1\hat{\mathbf{z}}\equiv \mathbf{h_1}$, 
and in ${\rm F_2}$
it can rotate in the $x$$-$$z$ plane, 
$\mathbf{h}=h_2\left(\sin\phi\hat{\mathbf{x}}+\cos\phi\hat{\mathbf{z}}\right)
\equiv \mathbf{h_2}$.
The single-particle Hamiltonian now reads\cite{cko}
${\cal H}_0=-({1}/{2m})({d^2}/{dy^2})+{\epsilon_{\perp}}-E_F(y)$, where 
$\epsilon_{\perp}\equiv k_\perp^2/2m$ denotes the transverse kinetic energy 
in the $x-z$ plane. Also,
$E_F(y)=E_{FS}\equiv{k_{FS}^2}/{2m}$ in the superconducting region and
$E_F(y)=E_{FM}\equiv{k_{FM}^2}/{2m}$
in the ferromagnetic layers. Throughout this paper,
we assume natural units $\hbar=k_B=1$ and measure all energies
in units of $E_{FS}$.
To take into account the more realistic situation
where the F materials can in general have different 
bandwidths than the S layer, 
we define (as in Ref.~\onlinecite{bv}) a  
mismatch parameter $\Lambda$
via $E_{FM}\equiv\Lambda E_{FS}$.%

We are aiming here to solve the problem in a fully self 
consistent manner. The self-consistent  pair potential $\Delta(y)$ can be expressed
in terms of the quasi-particle and quasi-hole wavefunctions. 
Accordingly,
\begin{equation}
\label{del}
\Delta(y) = \frac{g(y)}{2}{\sum_n}^\prime
\bigl[u_{n\uparrow}(y)v_{n\downarrow}^{\ast}(y)+
u_{n\downarrow}(y)v_{n\uparrow}^{\ast}(y)\bigr]\tanh\left(\frac{\epsilon_n}{2T}\right), \,
\end{equation}
where the primed sum is over 
all eigenstates with energies $\epsilon_n$
smaller than a characteristic Debye energy, and
$g(y)$ is 
the superconducting
coupling constant in the $S$ region and vanishes
elsewhere.
We obtain the 
self-consistent pair potential
by solving Eqs.~(\ref{bogo}) and (\ref{del})
following  the  iterative numerical procedures discussed
in  previous work.\cite{hv2p,cko}

\subsection{Application of the BTK method}
\label{BTK}
The  BTK formalism is a procedure to extract
the transmitted and reflected amplitudes, and hence the
conductance, from
solutions to the BdG equations.
This is accomplished by writing down the appropriate 
eigenfunctions in different regions. In this subsection,
we review the relevant aspects of the formalism\cite{btk}
for the non-self-consistent case (a step function pair
potential) with the objective of establishing notation
and methodology to describe,
in the next subsection, the procedure to be used in the
self-consistent case.

Consider first a spin-up quasi-particle
with energy $\epsilon$, incident
into the 
left
side labeled ``${\rm F}_1$", in Fig.~\ref{figure1}).
Since the exchange fields
in the ${\rm F_1}$  and ${\rm F_2}$  layers can be
non-collinear, it follows from Eq.~(\ref{bogo}) that
the spin-up (-down) quasi-particle wavefunction
is not just coupled to the spin-down (-up) quasi-hole
wavefunction, as is the case of F-S bilayers.
Indeed,
the wavefunction in the ${\rm F_1}$ layer is a linear combination of
the original incident spin-up quasi-particle wavefunctions
and various types of reflected wavefunctions, namely 
reflected spin-up and spin-down quasi-particle and
quasi-hole wavefunctions (via both ordinary
and Andreev reflections).
We use a single column vector notation to represent
these combinations,
\begin{equation}
\label{f1waveup}
\Psi_{F1,\uparrow}\equiv\begin{pmatrix}e^{ik^+_{\uparrow1}y}+b_{\uparrow}e^{-ik^+_{\uparrow1}y}
\\b_{\downarrow}e^{-ik^+_{\downarrow1}y}
\\a_{\uparrow}e^{ik^-_{\uparrow1}y}
\\a_{\downarrow}e^{ik^-_{\downarrow1}y}\end{pmatrix}\enspace\enspace.
\end{equation}
If the incident particle has spin down, the corresponding wavefunction
in ${\rm F_1}$ is
\begin{equation}
\label{f1wavedown}
\Psi_{F1,\downarrow}\equiv\begin{pmatrix}b_{\uparrow}e^{-ik^+_{\uparrow1}y}
\\e^{ik^+_{\downarrow1}y}+b_{\downarrow}e^{-ik^+_{\downarrow1}y}
\\a_{\uparrow}e^{ik^-_{\uparrow1}y}
\\a_{\downarrow}e^{ik^-_{\downarrow1}y}\end{pmatrix}\enspace\enspace.
\end{equation}
In these expressions $k^{\pm}_{\sigma 1}$
are quasi-particle $(+)$ and quasi-hole $(-)$ wavevectors
in the longitudinal direction $y$,
and satisfy the relation, 
\begin{equation}
\label{wavevector}
k^{\pm}_{\sigma m}=\left[\Lambda(1-\eta_{\sigma}{h}_m)\pm{\epsilon}-{k_\perp^2}\right]^{1/2},
\end{equation}
where $m=1$ (as used above) or $m=2$, used later. 
As mentioned above, all energies are in units of $E_{FS}$ and, in addition,
we measure all momenta in units of $k_{FS}$.
In this simple case, one can
easily distinguish the physical meaning
of each individual wavefunction. For instance in Eq.~(\ref{f1waveup}),
$a_{\downarrow}\left(0,0,0,1\right)^{T}e^{ik^-_{\downarrow1}y}$
is the reflected spin-down quasi-hole wavefunction.
The quasi-hole wavefunctions are the time reversed
solutions of the BdG equations and carry a positive sign in
the exponent for a left-going wavefunction. The relevant angles
can be easily found in terms of wavevector components. Thus, e.g., the
incident angle $\theta_i$ (for spin-up) at the ${\rm F_1-F_2}$ interface 
is
$\theta_{i}=\tan^{-1}\left({k_{\perp}}/{k_{\uparrow 1}^{+}}\right)$, 
and the Andreev reflected angle $\theta_{r\downarrow}^-$ for reflected
spin-down quasi-hole wavefunction is
$\theta_{r\downarrow}^-=\tan^{-1}\left({k_{\perp}}/{k_{\downarrow 1}^-}\right)$.
The conservation of transverse momentum leads to many important
features\cite{bv,zv1}
when one evaluates
the angularly averaged tunneling conductance, as we will see below.
For the intermediate layer ${\rm F_2}$,
the eigenfunction in general contains both
left- and right-moving plane waves, that is,
\begin{equation}
\label{f2wave}
\Psi_{F2}\equiv 
\begin{pmatrix}c_1f^+_{\uparrow}e^{ik^+_{\uparrow2}y}+c_2f^+_{\uparrow}e^{-ik^+_{\uparrow2}y}
+c_3g^+_{\uparrow}e^{ik^+_{\downarrow2}y}+c_4g^+_{\uparrow}e^{-ik^+_{\downarrow2}y}\\
c_1f^+_{\downarrow}e^{ik^+_{\uparrow2}y}+c_2f^+_{\downarrow}e^{-ik^+_{\uparrow2}y}
+c_3g^+_{\downarrow}e^{ik^+_{\downarrow2}y}+c_4g^+_{\downarrow}e^{-ik^+_{\downarrow2}y}\\
c_5f^-_{\uparrow}e^{ik^-_{\uparrow2}y}+c_6f^-_{\uparrow}e^{-ik^-_{\uparrow2}y}
+c_7g^-_{\uparrow}e^{ik^-_{\downarrow2}y}+c_8g^-_{\uparrow}e^{-ik^-_{\downarrow2}y}\\
c_5f^-_{\downarrow}e^{ik^-_{\uparrow2}y}+c_6f^-_{\downarrow}e^{-ik^-_{\uparrow2}y}
+c_7g^-_{\downarrow}e^{ik^-_{\downarrow2}y}+c_8g^-_{\downarrow}e^{-ik^-_{\downarrow2}y}
\end{pmatrix},
\end{equation}
where $k^{\pm}_{\uparrow 2}$ and $k^{\pm}_{\downarrow 2}$
are defined  in
Eq.~(\ref{wavevector}). 
The
$\pm$ indices are defined as previously, and the up and down arrows
refer to ${\rm F_1}$. The eigenspinors $f$ and $g$ that
correspond to spin parallel or antiparallel to ${\bf h_2}$ respectively,
are given, for
$0\leq\phi\leq{\pi}/{2}$, by the expression,
\begin{equation}
\label{f2spinor1}
\begin{pmatrix}f^+_{\uparrow}\\f^+_{\downarrow}\end{pmatrix}=
\frac{1}{\cal N}\begin{pmatrix}1\\\frac{1-\cos\phi}{\sin\phi}\end{pmatrix}
=\begin{pmatrix}f^-_{\uparrow}\\-f^-_{\downarrow}\end{pmatrix};\enspace
\begin{pmatrix}g^+_{\uparrow}\\g^+_{\downarrow}\end{pmatrix}=
\frac{1}{\cal N}\begin{pmatrix}-\frac{\sin\phi}{1+\cos\phi}\\1\end{pmatrix}
=\begin{pmatrix}-g^-_{\uparrow}\\g^-_{\downarrow}\end{pmatrix}
\end{equation}
with the normalization constant ${\cal N}=\sqrt{{2}/{1+\cos\phi}}$. 
These spinors reduce to those for pure spin-up and spin-down quasi-particles
and holes when $\phi=0$,
corresponding to a uniform magnetization along $z$. One can also easily 
see that the particular wavefunction
of Eq.~(\ref{f2wave}),
$c_1\left(f_{\uparrow}^+,f_{\downarrow}^+,0,0\right)^{T}e^{ik_{\uparrow 2}^+y}$
denotes a quasi-particle with spin parallel to the exchange field in ${\rm F_2}$.
When ${\pi}/{2}<\phi\leq\pi$,
these eigenspinors read
\begin{equation}
\label{f2spinor2}
\begin{pmatrix}f^+_{\uparrow}\\f^+_{\downarrow}\end{pmatrix}=
\frac{1}{\cal N}\begin{pmatrix}\frac{\sin\phi}{1-\cos\phi}\\1\end{pmatrix}
=\begin{pmatrix}-f^-_{\uparrow}\\f^-_{\downarrow}\end{pmatrix};\enspace
\begin{pmatrix}g^+_{\uparrow}\\g^+_{\downarrow}\end{pmatrix}=
\frac{1}{\cal N}\begin{pmatrix}1\\-\frac{1+\cos\phi}{\sin\phi}\end{pmatrix}
=\begin{pmatrix}g^-_{\uparrow}\\-g^-_{\downarrow}\end{pmatrix}
\end{equation}
with ${\cal N}=\sqrt{{2}/{1-\cos\phi}}$. 

In this subsection where we are still
assuming 
a non-self-consistent
stepwise potential equal to $\Delta_0$ throughout the 
S region and to zero elsewhere, we have 
the superconducting coherence factors, 
$\sqrt{2}u_0=\left[\left(\epsilon+{\sqrt{\epsilon^2-\Delta_0^2}}\right)/{\epsilon}\right]^{1/2}$ 
and $\sqrt{2}v_0=\left[\left(\epsilon-{\sqrt{\epsilon^2-\Delta_0^2}}\right)/{\epsilon}\right]^{1/2}$. 
In this case 
the right-going eigenfunctions
on the S side can be written as,
\begin{equation}
\Psi_S\equiv 
\begin{pmatrix}t_1u_0e^{ik^+y}+t_4v_0e^{-ik^-y}\\
t_2u_0e^{ik^+y}+t_3v_0e^{-ik^-y}\\
t_2v_0e^{ik^+y}+t_3u_0e^{-ik^-y}\\
t_1v_0e^{ik^+y}+t_4u_0e^{-ik^-y}
\end{pmatrix},
\label{transmitted}
\end{equation}
where, 
$k^{\pm}=\left[{1\pm\sqrt{{\epsilon}^2-{\Delta}_0^2}-k_\perp^2}\right]^{1/2}$ 
are quasi-particle (+) and quasi-hole (-) wavevectors in the S region. 
By using continuity of the four-component wavefunctions
and their first derivatives at both interfaces,
one can obtain all sixteen unknown coefficients in the above expressions
for the wavefunctions by solving
a set of linear equations of the
form ${\cal M}_{F1}x_{F1,\sigma}={\cal M}_{F2}x_{F2}$ 
at the ${\rm F_1-F_2}$ interface and $\tilde{{\cal M}}_{F2}x_{F2}={\cal M}_Sx_S$ 
at the ${\rm F_2-S}$ interface simultaneously,
where
\begin{subequations}
\label{interface}
\begin{align}
&x_{F1,\uparrow}^T=\left(1,b_{\uparrow},0,b_{\downarrow},0,a_{\uparrow},0,a_{\downarrow}\right)\\
&x_{F1,\downarrow}^T=\left(0,b_{\uparrow},1,b_{\downarrow},0,a_{\uparrow},0,a_{\downarrow}\right)\\ 
&x_{F2}^T=\left(c_1,c_2,c_3,c_4,c_5,c_6,c_7,c_8\right)\\
&x_{S}^T=\left(t_1,0,t_2,0,t_3,0,t_4,0\right),
\end{align}
\end{subequations}
and ${\cal M}_{F1}$, ${\cal M}_{F2}$, $\tilde{{\cal M}}_{F2}$,
and ${\cal M}_S$ are appropriate
$8\times8$ matrices, which are straightforward to write down. 
Use of these coefficients  gives us all the reflected and transmitted
amplitudes $a_\sigma$ and $b_\sigma$
which are used to compute the conductance, as discussed
in the next two subsections.

\subsection{Transfer matrix self consistent method}
\label{transfer}
The non-self-consistent step potential assumption is 
largely unrealistic. 
Proximity effects lead to a complicated oscillatory
behavior of the superconducting order parameter 
in the F layers and to the
generation\cite{Buzdin2005,Keizer2006,Bergeret2007,Wang2010,visani,Hubler2012,
Halterman2007,hv2p} of triplet pairs as discussed in Sec.~\ref{intro}. 
The concomitant depletion
of the pair amplitudes
near the F-S interface
means that
unless  the superconductor
is thick enough, the pair amplitude does not saturate
to its bulk value even deep inside the S regions.
Furthermore, as we shall emphasize below, lack of self
consistency may lead to violation of charge conservation: hence, 
while non-self-consistent approximations might be sometimes
adequate for equilibrium calculations, 
their use must be eschewed 
for transport. 
Therefore,
one should generally use a self-consistent pair potential 
that is allowed to spatially vary, as required
by  Eq.~(\ref{del}), and hence results in a minimum in the 
free energy of the system.

We begin by extending the BTK formalism 
to the spatially varying self-consistent pair potential obtained
as explained below Eq.~(\ref{del}).
Although the self-consistent solutions of the BdG equations
reveal that the pair amplitudes are non-zero
in the non-superconducting regions due to
the proximity effects, the pair potential vanishes in
these regions since $g(y)\equiv0$ there.
Therefore, one can still use Eqs.~(\ref{f1waveup}) and (\ref{f1wavedown}),
with
~(\ref{f2wave}), for the wavefunctions in the ${\rm F_1}$  and ${\rm F_2}$ regions.
To deal with the spatially varying pair potential on the S side,
we  divide it into many very thin layers 
with microscopic thicknesses of order $k_{FS}^{-1}$.
We treat each layer as a very thin superconductor
with a constant pair potential, $\Delta_i$, as
obtained from the self-consistent
procedure. We are then able to write the eigenfunctions 
of each superconducting layer corresponding to that
value of the pair potential.
For example, in the $i$-th layer, the eigenfunction
should contain all left and right going solutions, and it reads:
\begin{equation}
\label{sstate}
\Psi_{Si}\equiv 
\begin{pmatrix}
t_{1i}u_ie^{ik_i^+y}+\bar{t}_{1i}u_ie^{-ik_i^+y}
+t_{4i}v_ie^{-ik_i^-y}+\bar{t}_{4i}v_ie^{ik_i^-y}\\
t_{2i}u_ie^{ik_i^+y}+\bar{t}_{2i}u_ie^{-ik_i^+y}
+t_{3i}v_ie^{-ik_i^-y}+\bar{t}_{3i}v_ie^{ik_i^-y}\\
t_{2i}v_ie^{ik_i^+y}+\bar{t}_{2i}v_ie^{-ik_i^+y}
+t_{3i}u_ie^{-ik_i^-y}+\bar{t}_{3i}u_ie^{ik_i^-y}\\
t_{1i}v_ie^{ik_i^+y}+\bar{t}_{1i}v_ie^{-ik_i^+y}
+t_{4i}u_ie^{-ik_i^-y}+\bar{t}_{4i}v_ie^{ik_i^-y}
\end{pmatrix},
\end{equation}
where, 
$k_i^{\pm}=\left[1\pm\sqrt{{\epsilon}^2-{\Delta}_i^2}-k_{\perp}^2\right]^{1/2}$,
and ${\Delta}_i$ represents the strength of the
normalized self consistent pair potential in the
$i$-th superconducting layer. The superconducting coherence factors
$u_i$ and $v_i$  depend on $\Delta_i$ in the standard way. 
All the coefficients in
Eq.~(\ref{sstate}) are unknown, and remain to be determined.
However, in the outermost S layer (rightmost in our convention)
the eigenfunctions 
are of a form identical to Eq.~(\ref{transmitted})
but with different locally constant pair potential. 

We see then that the price one
has to pay  for including the proximity effects 
is the need to compute  a very large number of
coefficients. To do so, we  adopt here
a transfer matrix method to solve for these unknowns.\cite{strinati} If one
considers the interface between the $i$-th and
the $(i+1)$-th layer, we have the linear relation
$\tilde{{\cal M}}_ix_i={\cal M}_{i+1}x_{i+1}$, where, for a generic $i$, 
\begin{equation}
x_i^T=\left(t_{1i},\bar{t}_{1i},t_{2i},\bar{t}_{2i},
t_{3i},\bar{t}_{3i},t_{4i},\bar{t}_{4i}\right),
\end{equation}
and the matrices, $\tilde{{\cal M}}_i$ and ${\cal M}_{i+1}$, can be written 
as discussed in connection with Eq.~(\ref{interface}).
The coefficients in the $(i+1)$-th layer can be obtained in terms
of those in the $i$-th layer as
$x_{i+1}={\cal M}_{i+1}^{-1}\tilde{{\cal M}}_ix_i$.
In the same way, for the interface between the $(i-1)$-th layer
and the $i$-th layer, we can write
$x_i={\cal M}_i^{-1}\tilde{{\cal M}}_{i-1}x_{i-1}$. 
From the above relations, one can
write down the relation between $x_{i+1}$ and $x_{i-1}$,
i.e. $x_{i+1}={\cal M}_{i+1}^{-1}\tilde{{\cal M}}_i
{\cal M}_i^{-1}\tilde{{\cal M}}_{i-1}x_{i-1}$.
By iteration of this procedure, one can ``transfer'' the
coefficients
layer by layer and eventually relate the coefficients of the
rightmost layer, $x_n$, to those of the leftmost layer in S
and then on to the inner ferromagnetic layer ${\rm F_2}$:
\begin{equation} 
\label{transferring}
x_n={\cal M}_n^{-1}\tilde{{\cal M}}_{n-1}{\cal M}_{n-1}^{-1}
\cdots\tilde{{\cal M}}_1{\cal M}_1^{-1}
\tilde{{\cal M}}_{F2}x_{F2}
\end{equation}
By solving Eq.~(\ref{transferring}) together with
${\cal M}_{F1}x_{F1}={\cal M}_{F2}x_{F2}$, we obtain
all the coefficients in the ${\rm F_1}$ region, where the
wavefunction is {\it formally} still described 
by the expressions given in Eqs.~(\ref{f1waveup}) and
(\ref{f1wavedown}). Of course, all  coefficients involved, including
the energy dependent 
$a_\sigma$ and $b_\sigma$ values from which (see below)
the conductance is extracted,
are quite different from those in a non-self-consistent
calculation. These differences will be reflected in our results.
One can also prove that, when the pair potential in S is a constant
(non-self-consistent), then  ${\cal M}_{i+1}=\tilde{\cal M}_i$
and therefore Eq.~(\ref{transferring}) becomes
$x_n=x_1={\cal M}_1^{-1}\tilde{{\cal M}}_{F2}x_{F2}$. 
This  is formally identical to that we have seen
in our discussion of the non-self-consistent formalism.

This efficient technique, besides 
allowing us to determine all
the reflected and transmitted amplitudes in the outermost layers,
permits us to perform a consistency
check by recomputing the self-consistent solutions
to the BdG equations (the eigenfunctions).
Once we have determined the  amplitudes $x_{F1}$, $x_{F2}$, and $x_n$,  
we can use them to
find the amplitudes in any intermediate layer by
``transferring'' back the solutions. For example,
the coefficients $x_{n-1}$ can be found by using
$x_n={\cal M}_n^{-1}\tilde{\cal M}_{n-1}x_{n-1}$ 
if we know the coefficient $x_n$ for the rightmost layer.
Knowledge of these
coefficients in every region yields again 
the self-consistent
wavefunctions of the system. These of course should be the
same as the eigenfunctions found in the original procedure.
Although the numerical computations involved in this consistency
check are  rather intensive, it is worthwhile to perform them: 
we have verified that, by plugging these solutions into Eq.~(\ref{del})
and considering all possible solutions with all possible incident angles
to the BdG equations,
the output pair potential obtained from
the transport calculation is the same as the input pair potential
obtained by direct diagonalization. This would obviously not have  
been the case if the initial
pair  potential had not been fully self consistent to begin with.
The reflected and transmitted amplitudes calculated from the
self-consistent solutions are in general very different from 
the non-self-consistent ones and lead to 
different quantitative behavior of the tunneling conductance, as we shall 
discuss in section \ref{results}.

\subsection{Charge conservation}
\label{conservation}
We  discuss
now the  important
issue of the charge conservation laws. In  transport
calculations, it is fundamental to assure that they
are not violated \cite{baym}.  
From the Heisenberg
equation 
\begin{equation}
\frac{\partial}{\partial t}\left\langle\rho({\mathbf r})\right\rangle
=i\left\langle\left[{\cal H}_{eff},\rho({\mathbf r})\right]\right\rangle.
\end{equation}
By computing the above commutator,
we arrive at the following continuity condition
\begin{equation}
\frac{\partial}{\partial t}\left\langle\rho(\mathbf r)\right\rangle
+\nabla\cdot{\mathbf j}=
-4e{\rm Im}\left[\Delta({\mathbf r})\left\langle  
\psi_{\uparrow}^{\dagger}({\mathbf r})
\psi_{\downarrow}^{\dagger}({\mathbf r})\right\rangle\right].
\label{current}
\end{equation}
In the steady state, which is all that we are 
considering here, the first term on the left is 
omitted. Eqn.~(\ref{current}) is then simply an expression for
the divergence of the current. In our quasi one-dimensional 
system, and in terms of our wavefunctions, the conservation 
law can be rewritten as:
\begin{equation}
\frac{\partial j_y(y)}{\partial y}= 2e {\rm Im}\left\{\Delta(y)\sum_n\left[u_{n \uparrow}^* 
v_{n \downarrow}+u_{n\downarrow}^*v_{n\uparrow}\right]\tanh\left(\frac{\epsilon_n}{2T}\right)\right\}  
\label{currentuv}
\end{equation}
When the system is in equilibrium
the self-consistency condition
on the pair potential causes the right hand side of Eqs.~(\ref{current})
or~(\ref{currentuv}) 
to vanish. This would {\bf not} necessarily be the case
if a non-self-consistent\cite{imaginary}
solution were used.\cite{bagwell}
It was shown that charge conservation is only guaranteed 
when self consistency is adhered to in
microscopic Josephson junctions.\cite{sols2}
Current-voltage calculations for N-S heterostructures show that 
self-consistency is crucial to properly account for
all of the Andreev scattering channels arising when the current 
is constant throughout the system.\cite{sols}
While 
non-self-consistent solutions are less computationally 
demanding, 
their validity when calculating transport quantities 
in the nonequilibrium regime is always suspect. 
%

In the problem we are considering, there exists a finite 
voltage bias $V$
between the two leads of the system (see Fig.~\ref{figure1}).
This finite bias leads to
a non-equilibrium quasi-particle distribution and results
of course
in a net  current. Still, charge conservation 
must hold. To see how this works in this non-equilibrium case
we first write down the net quasi-particle charge density
in the $T\rightarrow 0$ limit (the case we consider 
here) by considering the excited state
$|\bf{k_1k_2}\cdots\rangle$ caused by the finite bias $V$.
Thus, this excited state contains all single particle states
$|\mathbf{k_j}\rangle$ ($j=1,2,\cdots$) 
with energies less than $eV$. For simplicity, let us first consider the
contribution by a single-particle state. We use $|\mathbf{k}\rangle$
to characterize this single particle state with an incident wavevector
$\mathbf{k}=\mathbf{k_{\perp}}+k\hat{\mathbf{y}}$ and energy
$\epsilon_\mathbf{k}$. The charge density associated with it is written as
\begin{align}
\rho&=-e\sum_{\sigma}\left\langle \mathbf{k} \left| \psi_{\sigma}^{\dagger}\psi_{\sigma} \right| \mathbf{k} \right\rangle \\\nonumber
&=-e\sum_{n\sigma}\left(|u_{n\sigma}|^2
\left\langle \mathbf{k} \left| \gamma_n^{\dagger}\gamma_n \right| \mathbf{k} \right\rangle
+|v_{n\sigma}|^2\left\langle \mathbf{k} \left| \gamma_n\gamma_n^{\dagger} \right| \mathbf{k} \right\rangle\right)\\\nonumber
&=-e\sum_{n\sigma}\left(|u_{n\sigma}|^2
\left\langle \mathbf{k} \left| \gamma_n^{\dagger}\gamma_n \right| \mathbf{k} \right\rangle
+|v_{n\sigma}|^2\left\langle \mathbf{k} \left| 1-\gamma_n^{\dagger}\gamma_n \right| \mathbf{k} \right\rangle\right)\\\nonumber
&=-e\sum_{n\sigma}|v_{n\sigma}|^2
-e\sum_{n\sigma}\left(|u_{n\sigma}|^2-|v_{n\sigma}|^2\right)\delta_{n\mathbf{k}}\\\nonumber
&=-e\sum_{n\sigma}|v_{n\sigma}|^2
-e\sum_{\sigma}\left(|u_{\mathbf{k}\sigma}|^2-|v_{\mathbf{k}\sigma}|^2\right)
\end{align}
The first term represents the ground state charge density.
For a generic excited state, $|\bf{k_1k_2}\cdots\rangle$ ,
that can contain many single-particle states,
one need to sum over all single-particle states for the charge
density such that
\begin{equation}
\rho=-e\sum_{n\sigma}|v_{n\sigma}|^2
-e\sum_{\epsilon_\mathbf{k}<eV}\sum_{\sigma}\left(|u_{\mathbf{k}\sigma}|^2-|v_{\mathbf{k}\sigma}|^2\right).
\end{equation}
The quasi-particle current density from this generic excited state
can also be computed,
\begin{align}
\label{current2}
j_y&=-\frac{e}{2m}\sum_{\epsilon_\mathbf{k}<eV}\sum_{\sigma} 
\left\langle-i\psi_{\sigma}^{\dagger}\frac{\partial}{\partial y}\psi_\sigma
+i\left(\frac{\partial}{\partial y}\psi_{\sigma}^{\dagger}\right)\psi_{\sigma}\right\rangle_{\mathbf{k}}\\\nonumber
&=-\frac{e}{m}{\rm Im}\left[\sum_{n\sigma}v_{n\sigma}\frac{\partial v_{n\sigma}^{\ast}}{\partial y}
+\sum_{\epsilon_\mathbf{k}<eV}\sum_{\sigma}\left(u_{\mathbf{k}\sigma}^{\ast}\frac{\partial u_{\mathbf{k}\sigma}}{\partial y}
+v_{\mathbf{k}\sigma}^{\ast}\frac{\partial v_{\mathbf{k}\sigma}}{\partial y}\right)\right]\\\nonumber
&=-\frac{e}{m}{\rm Im}\left[
\sum_{\epsilon_\mathbf{k}<eV}\sum_{\sigma}\left(u_{\mathbf{k}\sigma}^{\ast}\frac{\partial u_{\mathbf{k}\sigma}}{\partial y}
+v_{\mathbf{k}\sigma}^{\ast}\frac{\partial v_{\mathbf{k}\sigma}}{\partial y}\right)\right],
\end{align}
where $\langle...\rangle_\mathbf{k}$ is a shorthand notation of
$\langle\mathbf{k}\left|...\right|\mathbf{k}\rangle$.
The first term in the second line vanishes
because it represents
the net current for the system in the ground state
with a real pair potential.
The right hand side of the continuity equation, Eq.~(\ref{currentuv}),
becomes
$-4e{\rm Im}\left[\Delta\sum_{\epsilon_\mathbf{k}<eV}\left(u_{\mathbf{k}\uparrow}^{\ast}
v_{\mathbf{k}\downarrow}+v_{\mathbf{k}\uparrow}u_{\mathbf{k}\downarrow}^{\ast}\right)\right]$
and is responsible for the interchange
between the quasi-particle current density and
the supercurrent density\cite{btk}.
We have numerically verified that by properly
including  these  terms,
all of our numerical results for the current density
are constant throughout the whole system.

\subsection{Extraction of the conductance}
\label{extraction}
We are now in a position to compute the
differential tunneling conductances.
We begin  by  discussing the extraction 
of the conductance from the BTK theory.
As we mentioned in the previous subsection,
the finite bias $V$ and the resulting
non-equilibrium distribution leads to an electric current
flowing in the junction.
In the BTK theory, this current can be evaluated from the
following\cite{btk} expression,
\begin{equation}
\label{totalcurrent}
I(V)=\int G(\epsilon)\left[f\left(\epsilon-eV\right)
-f\left(\epsilon\right)\right]d\epsilon,
\end{equation}
where $f$ is the Fermi function. The
energy dependent tunneling conductance,
$G(\epsilon)=\partial I/{\partial V}|_{V=\epsilon}$ in the low-$T$ limit, is given as:
\begin{align}
\label{conductance}
&G(\epsilon,\theta_i)=\sum_\sigma P_\sigma G_{\sigma }(\epsilon,\theta_i) 
\\\nonumber
&=\sum_{\sigma}P_{\sigma}\left(1+\frac{k^-_{\uparrow 1}}{k^+_{\sigma 1}}|a_{\uparrow }|^2
+\frac{k^-_{\downarrow 1}}{k^+_{\sigma 1}}|a_{\downarrow }|^2
-\frac{k^+_{\uparrow 1}}{k^+_{\sigma 1}}|b_{\uparrow }|^2
-\frac{k^+_{\downarrow 1}}{k^+_{\sigma 1}}|b_{\downarrow }|^2\right), 
\end{align} 
where we have used, as is customary,
natural units of conductance $(e^2/h)$. In the
above expression the different $k$ components are as
explained in subsection \ref{BTK} (see e.g. Eq.~(\ref{wavevector}))
and the $a_\sigma$ and $b_\sigma$
are as defined in Eqns.~(\ref{f1waveup}) 
and (\ref{f1wavedown}). 
These coefficients, which are of course energy 
dependent, are calculated using the self-consistent
transfer matrix technique of subsection \ref{transfer}. Therefore,
even though Eq.~(\ref{conductance}) is formally the same in the self-consistent
and non-self-consistent cases, the results for the reflection amplitudes
or probabilities involved,
$|a_{\uparrow}|^2$, $|a_{\downarrow}|^2$, $|b_{\uparrow}|^2$,
and $|b_{\downarrow}|^2$ are different in these two schemes. 
The angle $\theta_i$ is the incident angle, discussed in terms
of ${\bf k}$ components below Eq.~(\ref{wavevector}).
The weight factor $P_{\sigma}\equiv\left(1-h_1\eta_{\sigma}\right)/2$
accounts for the number of available states for spin-up
and spin-down bands in the outer electrode.
The tunneling conductance can also be interpreted as the
transmission coefficient for electrical current.
The method enables us also to compute the current density directly from the
wavefunctions, Eqs.~(\ref{f1waveup}) and
(\ref{f1wavedown}),
in the ${\rm F_1}$ layer by using Eq.~(\ref{current2}) and we have
been able to verify that the resulting current density
is identical to the terms inside the bracket
in the expression of $G(\epsilon)$, Eq.~(\ref{conductance}).
In other words, in the low-$T$ limit
the continuum-limit version of Eq.~(\ref{current2})
is equivalent to Eq.~(\ref{totalcurrent}). 

The conductance results Eq.~(\ref{conductance})
also depend on the incident angle of electrons, $\theta_i$.
Experimentally, one can measure the forward conductance, $\theta_i=0$,
via point contacts or, in most other experimental conditions,
an angular average. Consequently, it is worthwhile
to compute the angularly averaged conductance by
using the following definitions,
\begin{equation}
\label{aaG}
\langle G_{\sigma}({\epsilon})
\rangle=\frac{\int_0^{\theta_{c\sigma}} d\theta_i \cos\theta_i G_\sigma(\epsilon,\theta_i)}
{\int_0^{\theta_{c\sigma}} d\theta_i \cos\theta_i},
\end{equation}
and
\begin{equation}
\langle G \rangle=\sum_{\sigma} P_{\sigma}\langle G_{\sigma}\rangle,
\end{equation}
where the critical angle $\theta_{c\sigma}$ is in general different for
spin-up and spin-down bands.
This 
critical angle arises from the conservation
of transverse momentum and the corresponding Snell law:
\begin{equation}
\label{snell}
\begin{aligned}
&\sqrt{\left({k_{\sigma 1}^+}^2+k_{\perp}^2\right)}\sin\theta_i
=\sqrt{\left({k_{\sigma' 1}^+}^2+k_{\perp}^2\right)}\sin\theta_{r\sigma'}^+\\
&=\sqrt{\left({k_{\sigma' 1}^-}^2+k_{\perp}^2\right)}\sin\theta_{r\sigma'}^-
=\sin\theta_S,
\end{aligned}
\end{equation}
where we continue to measure wavevectors in units
of $k_{FS}$. The angles $\theta_{r\sigma}^\pm$ 
satisfy $\tan^{-1}\left({k_{\perp}}/{k_{\sigma 1}^{\pm}}\right)$, 
and the $\sigma$ and $\sigma'$ are each  $\uparrow$ 
or $\downarrow$. The last equality
in Eq.~(\ref{snell}) represents the case of
the transmitted wave in S, and $\theta_S$ is
the transmitted angle.
Although the self-consistent pair potential varies in S and
so do the quasi-particle (hole) wavevectors, we here need only
consider the transmitted angle $\theta_S$ in the rightmost
layer: this follows in the same way
as the usual Snell's law 
in a layered system, as given in 
elementary textbooks.
From Eq.~(\ref{snell}), one can determine the critical angles
for different channels.
Consider, e.g.,
a spin-up electron incident from  ${\rm F_1}$  without
any Fermi wavevector mismatch, i.e. $\Lambda=1$.
Since we are only concerned with the case that
the bias of tunneling junctions is of the order
of superconducting gap and therefore much smaller
than the Fermi energy, the approximate magnitude
of the incident wavevector is $\sqrt{1+{h}_1}$, 
the Andreev approximation. 
We substitute this and similar expressions into Eq.~(\ref{snell})  and,
with the help of Eq.~(\ref{wavevector}), we obtain
\begin{equation}
\label{andreevr}
\sqrt{1+{h}_1}\sin\theta_i=\sqrt{1-{h}_1}\sin\theta_{r\downarrow}^-
=\sin\theta_S.
\end{equation}
One can straightforwardly verify that, when the relation
$\theta_i>\sin^{-1}[((1-{h}_1)/(1+{h}_1))^{1/2}]$ 
is satisfied for the incident angle, the conventional Andreev reflection
becomes an evanescent wave\cite{zv2}.
In this case, the conventional Andreev reflection does not contribute
to the angular averaging.
On the other hand, if the energy $\epsilon$ of
the incident electron is less than the saturated value of 
the superconducting pair amplitude in S,
all the contribution to the conductance from
the transmitted waves in S also vanishes because
$k^{\pm}$ acquires an imaginary part.
However, even the condition that
$\epsilon$ is greater than the saturated superconducting 
amplitude does not guarantee that the contribution from
the transmitted waves to
the conductance is nonvanishing.
One still needs to consider the transmitted critical angle
$\sin^{-1}[1/(1+h_1)^{1/2}]$. 
We define the critical angle $\theta_{c\sigma}$ to be the largest one
among all the reflected and transmitted critical angles.
It is obvious that the critical angles $\theta_{c\sigma}$
are different for spin-up and spin-down bands when $h_1\neq0$. 

\subsection{Spin transport}
\label{spintran} 

We consider
now the spin-transfer torque 
and the spin current.
As the charge carriers that flow through
our system, along the 
$y$ direction in our convention, are spin polarized,
the STT provides an additional probe of the spin degree of freedom. 
Unlike the charge current, that must be a constant throughout
the system, the spin current density is generally not a conserved
quantity in the ferromagnet regions as we will demonstrate below. 
The discussion in Sec.~\ref{conservation} on how the BTK formalism deals with
the charge current can be extended
to compute these spin dependent transport quantities.
We need here the continuity equation for the local magnetization 
$\mathbf{m}\equiv-\mu_B\sum_{\sigma}\left\langle\psi^{\dagger}_{\sigma} 
\bm{\sigma}\psi_{\sigma}\right\rangle$, where $\mu_B$ is the Bohr magneton.
By using the Heisenberg equation 
$\frac{\partial}{\partial t}\left\langle\mathbf{m}({\mathbf r})\right\rangle
=i\left\langle\left[{\cal H}_{eff},\mathbf{m}({\mathbf r})\right]\right\rangle$
we obtain the relation: 
\begin{equation}
\label{spinconserve}
\frac{\partial}{\partial t}\langle m_i \rangle+ \frac{\partial}{\partial y} S_i= \tau_i,
\enspace\enspace i=x,y,z 
\end{equation}
where  $\bm{\tau}$ is the spin-transfer torque,
$\bm{\tau}\equiv 2\mathbf{m}\times\mathbf{h}$, 
 and
the spin current density $S_i$ is given by
\begin{equation}
S_i\equiv\frac{i\mu_B}{2m}\sum_\sigma\left\langle \psi_\sigma^{\dagger}\sigma_i\frac{\partial \psi_\sigma}{\partial y}
-\frac{\partial \psi_\sigma^{\dagger}}{\partial y}\sigma_i\psi_\sigma\right\rangle.
\end{equation}
The spin current density 
reduces from a tensor form to a vector  because of the  
quasi-one-dimensional nature of our geometry.
From Eq.~(\ref{spinconserve}), we can see that $\mathbf{S}$ 
is a local physical quantity and $\bm{\tau}$ is
responsible for the change of local magnetizations due
to the flow of spin-polarized currents.
As we shall see in Sec.~\ref{results},
the conservation law (with the source torque term) 
 for the spin density is fundamental
and one has to  check it is not violated when  
studying these transport quantities. 

In the low-$T$ limit and with the presence of a finite bias, the non-equilibrium local magnetizations
$m_i\equiv\sum_{\epsilon_\mathbf{k}<eV}\sum_{\sigma}-\mu_B\langle\psi_\sigma^\dagger\sigma_i\psi_\sigma\rangle_\mathbf{k}$
in Eq.~(\ref{spinconserve}) reads 
\begin{subequations}
\label{mag} 
\begin{align}
m_x=&-\mu_B\left[\sum_n\left(-v_{n\uparrow}v_{n\downarrow}^{\ast}-v_{n\downarrow}v_{n\uparrow}^{\ast}\right)\right.\\\nonumber
&\left.+\sum_{\epsilon_\mathbf{k}<eV}\left(u_{\mathbf{k}\uparrow}^{\ast}u_{\mathbf{k}\downarrow}
+v_{\mathbf{k}\uparrow}v_{\mathbf{k}\downarrow}^{\ast}
+u_{\mathbf{k}\downarrow}^{\ast}u_{\mathbf{k}\uparrow}
+v_{\mathbf{k}\downarrow}v_{\mathbf{k}\uparrow}^{\ast}\right)\right]\\
m_y=&-\mu_B\left[i\sum_n\left(v_{n\uparrow}v_{n\downarrow}^{\ast}-v_{n\downarrow}v_{n\uparrow}^{\ast}\right)\right.\\\nonumber 
&\left.-i\sum_{\epsilon_\mathbf{k}<eV}\left(u_{\mathbf{k}\uparrow}^{\ast}u_{\mathbf{k}\downarrow}
+v_{\mathbf{k}\uparrow}v_{\mathbf{k}\downarrow}^{\ast}
-u_{\mathbf{k}\downarrow}^{\ast}u_{\mathbf{k}\uparrow}
-v_{\mathbf{k}\downarrow}v_{\mathbf{k}\uparrow}^{\ast}\right)\right]\\
m_z=&-\mu_B\left[\sum_n\left(|v_{n\uparrow}|^2-|v_{n\downarrow}|^2\right)\right.\\\nonumber
&\left.+\sum_{\epsilon_\mathbf{k}<eV}\left(|u_{\mathbf{k}\uparrow}|^2
-|v_{\mathbf{k}\uparrow}|^2
-|u_{\mathbf{k}\downarrow}|^2
+|v_{\mathbf{k}\downarrow}|^2\right)\right],
\end{align}
\end{subequations}
where the first summations in the expressions for $m_i$ denote the ground state
local magnetizations. The second summations appear
as a consequence of the finite bias between electrodes.
The expressions for the corresponding spin currents,
\begin{equation}
S_i\equiv\frac{i\mu_B}{2m}\sum_{\epsilon_\mathbf{k}<eV}\sum_{\sigma}\left\langle \psi_\sigma^{\dagger}\sigma_i\frac{\partial \psi_\sigma}{\partial y}
-\frac{\partial \psi_\sigma^{\dagger}}{\partial y}\sigma_i\psi_\sigma\right\rangle_\mathbf{k},
\end{equation}
becomes
\begin{subequations}
\label{spincur}
\begin{align}
S_x=&\frac{-\mu_B}{m}{\rm Im}\left[\sum_n\left
(-v_{n\uparrow}\frac{\partial v_{n\downarrow}^{\ast}}{\partial y}
-v_{n\downarrow}\frac{\partial v_{n\uparrow}^{\ast}}{\partial y}\right)\right.\\\nonumber
&\left.+\sum_{\epsilon_\mathbf{k}<eV}\left(u_{\mathbf{k}\uparrow}^{\ast}\frac{\partial u_{\mathbf{k}\downarrow}}{\partial y}
+v_{\mathbf{k}\uparrow}\frac{\partial v_{\mathbf{k}\downarrow}^{\ast}}{\partial y}
+u_{\mathbf{k}\downarrow}^{\ast}\frac{\partial u_{\mathbf{k}\uparrow}}{\partial y}
+v_{\mathbf{k}\downarrow}\frac{\partial v_{\mathbf{k}\uparrow}^{\ast}}{\partial y}\right)\right]\\
S_y=&\frac{\mu_B}{m}{\rm Re}\left[\sum_n\left(-v_{n\uparrow}
\frac{\partial v_{n\downarrow}^{\ast}}{\partial y}
+v_{n\downarrow}\frac{\partial v_{n\uparrow}^{\ast}}{\partial y}\right)\right.\\\nonumber
&\left.+\sum_{\epsilon_\mathbf{k}<eV}\left(u_{\mathbf{k}\uparrow}^{\ast}\frac{\partial u_{\mathbf{k}\downarrow}}{\partial y}
+v_{\mathbf{k}\uparrow}\frac{\partial v_{\mathbf{k}\downarrow}^{\ast}}{\partial y}
-u_{\mathbf{k}\downarrow}^{\ast}\frac{\partial u_{\mathbf{k}\uparrow}}{\partial y}
-v_{\mathbf{k}\downarrow}\frac{\partial v_{\mathbf{k}\uparrow}^{\ast}}{\partial y}\right)\right]\\
S_z=&\frac{-\mu_B}{m}{\rm Im}\left[\sum_n\left(v_{n\uparrow}\frac{\partial v_{n\uparrow}^{\ast}}{\partial y}
-v_{n\downarrow}\frac{\partial v_{n\downarrow}^{\ast}}{\partial y}\right)\right.\\\nonumber
&\left.+\sum_{\epsilon_\mathbf{k}<eV}\left(u_{\mathbf{k}\uparrow}^{\ast}\frac{\partial u_{\mathbf{k}\uparrow}}{\partial y}
-v_{\mathbf{k}\uparrow}\frac{\partial v_{\mathbf{k}\uparrow}^{\ast}}{\partial y}
-u_{\mathbf{k}\downarrow}^{\ast}\frac{\partial u_{\mathbf{k}\downarrow}}{\partial y}
+v_{\mathbf{k}\downarrow}\frac{\partial
v_{\mathbf{k}\downarrow}^{\ast}}{\partial y}\right)\right]. 
\end{align}
\end{subequations}
The first summations in Eq.~(\ref{spincur}) represent 
the static spin current densities when there is no bias.
The static spin current does not need to vanish, since
a static spin-transfer torque may exist near the
boundary of two magnets with misaligned exchange fields.
The finite bias leads to a non-equilibrium quasi-particle
distribution for the system and results in 
non-static spin current densities that are represented
by the second summation in Eq.~\ref{spincur}.
Obviously, the spin-transfer torque has to vanish
in the superconductor where the exchange field is
zero. 
It is conventional to normalize $\mathbf{m}$ to\cite{Halterman2007}
$-\mu_B(N_\uparrow+N_\downarrow)$, where the number densities
$N_\uparrow=k_{FS}^3(1+h_m)^{3/2}/(6\pi^2)$ and 
$N_\downarrow=k_{FS}^3(1-h_m)^{3/2}/(6\pi^2)$. 
Following this convention, we normalize $\bm{\tau}$ to
$-\mu_B(N_\uparrow+N_\downarrow)E_{FS}$ and $\mathbf{S}$
to $-\mu_B(N_\uparrow+N_\downarrow)E_{FS}/k_{FS}$.



\section{Results}
\label{results}

\begin{figure} 
\includegraphics[width=0.45\textwidth] {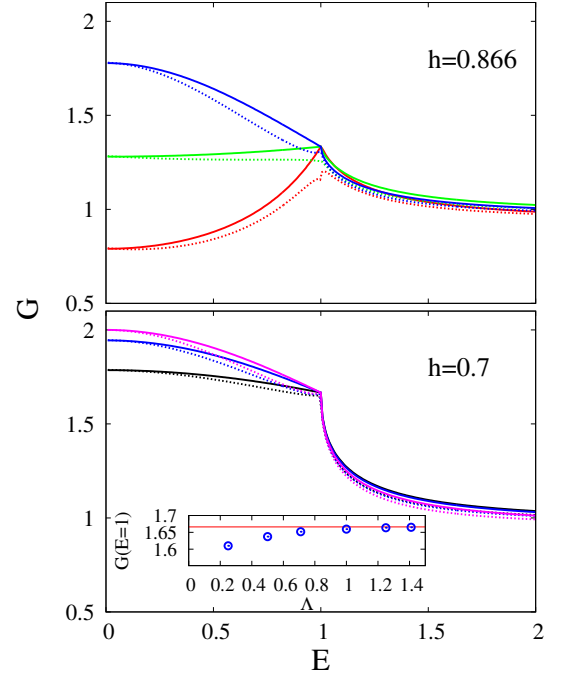} 
\caption{(Color online) 
Bias dependence of the results for the 
forward conductance, $G$, in
thick F-S bilayers (see text). The values of $h$ are indicated. 
In both main panels the
solid and dashed curves show $G$, in units
of $e^2/h$ for non-self-consistent and self-consistent
results, respectively. The bias $E$ is in units
of the S bulk gap $\Delta_0$. In the top panel the (red) lower
curves are for a mismatch parameter $\Lambda=0.25$,  (green)
the middle  curves  for $\Lambda=0.5$, and the (blue)
higher curves for $\Lambda=1$. In the bottom panel, the (purple) 
top curves are for $\Lambda=1.41$, the (blue) curves are as in the top panel, 
and the (black) lower ones for $\Lambda=0.71$. The inset (see text) shows 
$G(E=1)$ vs. $\Lambda$ in the self consistent calculation (dots) 
and the non-self-consistent result (line).
}
\label{figure2}
\end{figure}

The forward scattering conductances  $G$ 
are computed by considering
a particle incident with an angle $\theta_i\cong0$ 
(normal incidence). Angular averaging has been discussed 
in the text above Eq.~(\ref{aaG}). 
The bias energy $E\equiv eV$ is in units of the zero
temperature gap, $\Delta_0$, in bulk S material and $e^2/h$ is
used as the natural unit of conductance. 
When the ${\rm F_1}$  and ${\rm F_2}$ regions are made of same F material,
i.e., $h_1=h_2$ and $k_{\uparrow1,(\downarrow1)}^{\pm}=k_{\uparrow2,(\downarrow)2}^{\pm}$, 
we will use $h$ (not to be confused with Planck's constant)
and  $k_{\uparrow,(\downarrow)}^{\pm}$ to denote their exchange 
fields and wavevectors. This is the case we will mostly study.
All results are for the low-$T$ limit. 
All of the lengths 
are measured in unit of $k_{FS}^{-1}$ and denoted by capital letters,
e.g. $D_S$ denotes $k_{FS} d_S$.

\subsection{Bilayers}
\label{bilayers}
We begin with a brief discussion of self-consistent results for the 
tunneling conductance in F-S bilayers, contrasting them with 
non-self-consistent results. We assume that the S layer 
is very thick so that the pair amplitude saturates to its bulk value 
deep inside the S region. In this subsection,
the dimensionless superconducting coherence length $\Xi_0$ is taken to be
$50$ and the thicknesses $D_F$ and $D_S$ of the
F and S layers are both $15\Xi_0$.
By computing the pair amplitudes via the direct diagonalization
method,\cite{hv2p} we have verified that they  
indeed saturate to their bulk value 
with this large  ratio of $D_S$ to $\Xi_0$.

As discussed in Sec.~\ref{intro}, 
the replacement of non-magnetic metals with ferromagnets 
in a bilayer leads to strong suppression 
of the Andreev reflection in the subgap region.
The decrease of the 
zero bias conductance (ZBC) strongly depends on the magnitude 
of the exchange field in F. This dependence is used to measure 
the degree of spin-polarization of magnetic materials
experimentally.\cite{soulen,upad} 
However, in early theoretical work,\cite{zv1,zv2} it was shown that 
to  
accurately determine the degree of spin-polarization, one has to 
consider the Fermi wavevector mismatch (FWM), $\Lambda$, as well as the 
interfacial barriers. The ZBC peak
is very sensitive to both spin-polarization and  FWM and the 
dependence cannot be characterized by a single parameter.

We display in Fig.~\ref{figure2} forward
conductance vs. bias results for both the self-consistent and  
non-self-consistent calculations, 
at two different 
values of the exchange fields
and several FWM values.  
One sees at once that the self-consistent results approach  the 
non-self-consistent ones in the zero bias limit, while 
deviating the most for energies near the superconducting gap. 
The ZBC decreases with increasing $h$ and with decreasing $\Lambda$.
Also, larger $h$ indeed leads to a conspicuous reduction in
the subgap conductance  and so does 
the introduction of FWM. 
One 
can conclude that the behavior of the ZBC can not be characterized 
by only one parameter, either $h$ or $\Lambda$. Instead, one should
expand the fitting parameter space to determine the degree of 
spin polarization.

In the non-self-consistent framework, 
the conductance at the superconducting gap ($E=1$ in
our units) is independent of 
$\Lambda$ at a given $h$. However,  
earlier work\cite{bv} predicted that  
this conclusion is invalid in self-consistent approach,
and that the conductance at the
superconducting gap varies monotonically with increasing $\Lambda$.
Here we verify this via our self-consistent transfer matrix method. 
The inset in the bottom panel 
of Fig.~\ref{figure2} clearly shows this dependence on $\Lambda$. 
Figure~\ref{figure2} also shows that the self-consistent results 
(dashed curves) on subgap conductances are in general lower than 
those obtained in the non-self-consistent framework (solid curves) 
for a strong exchange field. 
On the other hand, in the high bias limit, the 
self-consistent results become similar to the non-self-consistent ones. 
This is simply because the particle does not experience much of a difference  
between a step-like pair potential and a smooth pair potential
when it is incident with high enough energy.
Finally,  clear cusps appear
at the superconducting gap value in some cases, e.g., the forward scattering
conductance curve at $h=0.866$ and $\Lambda=1$. 
This is consistent with what is found
in previous work\cite{bv} for thick bilayers. 
\begin{figure} 
\includegraphics[width=0.45\textwidth] {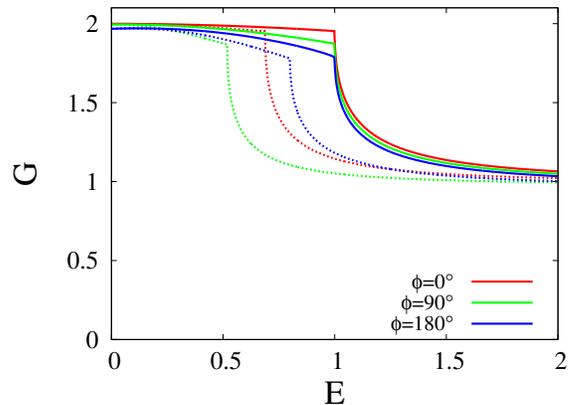} 
\caption{(Color online) Comparison between the 
self-consistent and non-self-consistent forward scattering conductances of
${\rm F_1F_2S}$ trilayers. 
The solid and the dashed lines are for non-self-consistent
and self-consistent results respectively. The (red) curves,
highest at the critical bias (CB) 
are  for $\phi=0^{\circ}$. The (blue) curves, lowest at CB,
are for $\phi=180^{\circ}$.
We have $D_{F1}=10$, $D_{F2}=12$, and $D_S=180$ (see text).} 
\label{figure3}
\end{figure}
\subsection{Trilayers}
\label{trilayers}
We now  discuss our results for 
${\rm F_1F_2S}$ trilayers of finite widths. First, we discuss 
the dependence of the tunneling conductances
on the angle $\phi$ between ${\mathbf h_1}$ and ${\mathbf h_2}$ (see
below Eq.~(\ref{bogo}) and Fig.~\ref{figure1}). 
An important reason for  
considering trilayers with finite widths is 
the strong dependence of the superconducting transition temperatures $T_c$ 
on the angle $\phi$ 
due to  proximity effects\cite{cko} 
and induced long-range triplet correlations.\cite{gol}
Field induced switching effects\cite{oh} 
also make these structures attractive candidates for memory elements.
The  non-monotonic behavior of $T_c(\phi)$ 
with its minimum  being near
$\phi=90^{\circ}$,
was extensively discussed
in Ref.~\onlinecite{cko}. 
This angular dependence  has been shown to be 
related to the induced triplet pairing correlations\cite{alejandro}.
The superconducting
transition temperatures are also 
predicted to be positively correlated with the 
singlet pair amplitudes deep inside the S regions\cite{cko}. 
Therefore, it is of particular importance 
to consider systems of finite size to take into view the whole
picture of proximity effects
on the angular dependence of the tunneling conductance. 
For the results shown in this subsection, 
we assume the absence of FWM ($\Lambda=1$).


\begin{figure*}
\includegraphics[width=0.9\textwidth] {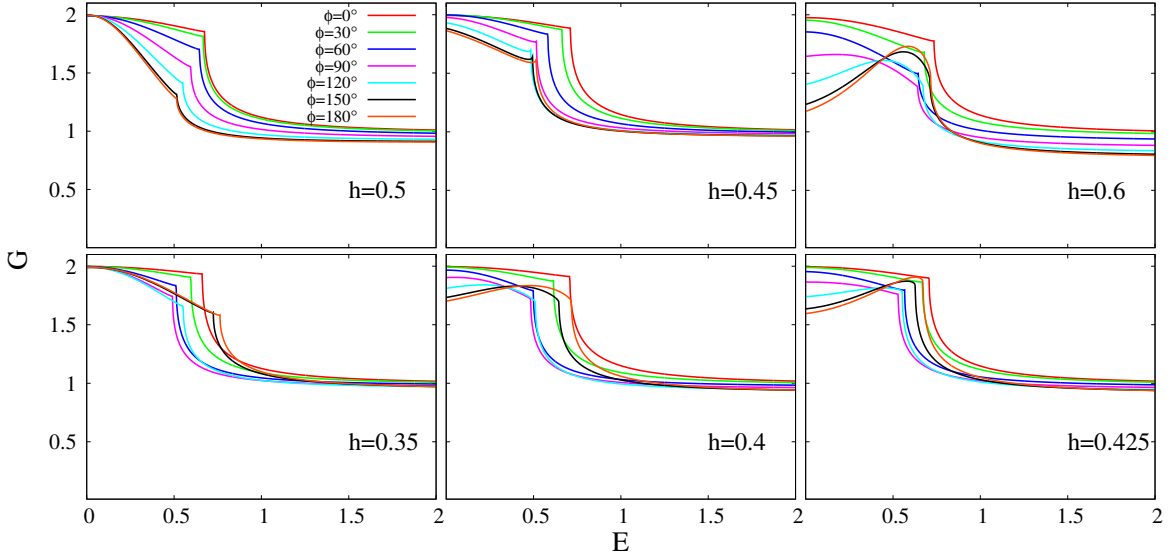} 
\caption{(Color online) Forward scattering conductance of ${\rm F_1F_2S}$ 
trilayers for several angles $\phi$ as indicated in the legend.
The top panels are for $D_{F1}=10$, $D_{F2}=12$, and $D_{S}=180$ 
and the bottom panels for $D_{F1}=10$, $D_{F2}=18$, and $D_{S}=180$. 
The  exchange field strength $h$ is indicated.
For the left panels, the conductances at CB decrease with increasing $\phi$.
For the other panels, the ZBC (see text) decreases as $\phi$ increases. }
\label{figure45}
\end{figure*}

\subsubsection{Forward Scattering}
As a typical example of our
results, we show in Fig.~\ref{figure3} results for
the $\phi$ 
dependence of the forward 
scattering conductances. 
The exchange field we use here for both F 
layers 
is $h=0.3$, 
and the thicknesses of
the F$_1$ and F$_2$ layers correspond to $D_{F1}=10$ and $D_{F2}=12$ respectively, 
while the S layer has  width $D_S=180=1.5 \Xi_0$.
Results obtained via the non-self-consistent approach are 
plotted for comparison.
In the non-self-consistent framework where  
the single parameter $\Delta_0$ describes the stepwise pair potential, 
one sees in Fig.~\ref{figure3} that for all 
values of the angle $\phi$ 
the conductance curves drop when the bias is at
$\Delta_0$, corresponding to $E=1$ in our units. 
In contrast, for the self-consistent results, one can clearly see in
Fig.~\ref{figure3},
that 
the drop in the conductance curves occurs at different bias values 
for different angles. We
also see that this critical bias (which
we will denote by CB) depends on $\phi$ non-monotonically,
with 
$\phi=180^{\circ}$ corresponding to the largest and $\phi=90^{\circ}$
to the smallest bias values. 
Since the CB depends on the strength of 
the superconducting gap deep inside
the S regions, the non-monotonicity of the CB in Fig.~\ref{figure3} 
is  correlated 
with the non-monotonicity of $T_c$. 
The CB never reaches 
unity, 
in these trilayers, due to their finite size.
Accordingly, this feature of the correct self-consistent results implies
that one cannot adequately determine the  angular
dependence of the 
forward conductance in the non-self-consistent framework.
This feature also provides experimentalists 
with another way to measure the strength
of the superconducting gap for different angles in these trilayers by
determining the CB in a set of conductance curves.
The remaining results shown in this section are all computed 
self-consistently. 

In Fig.~\ref{figure45}, we present more results for 
the dependence of the forward scattering conductances
on $\phi$. 
In the top panels 
the thicknesses of 
each layer and the coherence length are the same as Fig.~\ref{figure3}. 
In the bottom panels 
we increase the thickness 
of the inner magnetic layer to $D_{F2}=18$ while $D_{F1}$, $D_{S}$, 
and $\Xi_0$ remain unchanged. 
For each row of Fig.~\ref{figure45}, results for three different exchange 
fields are plotted. 
In the top left panel ($h=0.5$) 
we see that the angular dependence 
of the CB (or the magnitude of the saturated pair amplitudes) 
is monotonic with
$\phi$. Although this monotonicity
is not common, we have verified that it 
is  consistent with the theoretical results for $T_c(\phi)$ 
for the same particular case.
The more usual non-monotonic dependence is  found 
in 
all other panels, as discussed 
in the previous paragraph. In every case, we have also 
checked that the magnitude of the CB reflects the magnitude of the 
self-consistent pair amplitudes deep inside the superconductor.

For the ZBC, we see that the degree of its 
angular dependence is very sensitive to  $h$. 
In the top left panel, with $h=0.5$, the ZBC 
is nearly independent on $\phi$. On the other hand, the ZBC in the 
top right panel,  $h=0.6$,  drops 
by almost a factor of two as
$\phi$ varies from the relative parallel (P) orientation, $\phi=0^{\circ}$, 
to the antiparallel (AP) orientation, $\phi=180^{\circ}$. 
This is a consequence of interference
between the spin-up and spin-down wavefunctions
under the influence of the rotated exchange field 
in the middle layer. 
In the top left panel, 
we see that the conductance at CB decreases with increasing angle.
In other words, the zero bias conductance peak (ZBCP) becomes
more prominent as $\phi$ is increased. However,  
for the top middle panel,  $h=0.45$, the development of the ZBCP 
is less noticeable when the angle $\phi$ is increased. 
In the top right panel, $h=0.6$, the ZBCP evolves into a 
zero bias conductance dip (ZBCD) as $\phi$ varies
from $\phi=0^{\circ}$ to 
$\phi=180^{\circ}$, with a clear finite bias conductance peak (FBCP)
appearing just below the CB. This behavior is reminiscent\cite{zv2} of that
which occurs when a barrier, or mismatch, are present.
In the bottom panels of this figure, corresponding to 
a larger value of $D_{F2}$ 
one can observe similar features. 
For example,  
a slight change from $h=0.35$ to $h=0.4$ causes by itself a very large 
change in the behavior of the
ZBC. Moreover, the evolution of the ZBCP to a ZBCD 
accompanies the occurrence of a
FBCP when $\phi>90^{\circ}$. The location of the FBCP
also moves closer to the CB value when $\phi$ increases.
That these features of the ZBC depend on both the strength of exchange
field (reflected in $k_{\uparrow}^{\pm}$ and $k_{\downarrow}^{\pm}$) and 
the thickness of the ${\rm F_2}$ layer indicates that the ZBC shows 
the characteristics of a resonance scattering
phenomenon as in an elementary quantum mechanical barrier.
The main difference is that the scattering problem here involves
the intricate interference between quasi-particle and 
quasi-hole spinors. 

\begin{figure}
\includegraphics[width=0.45\textwidth] {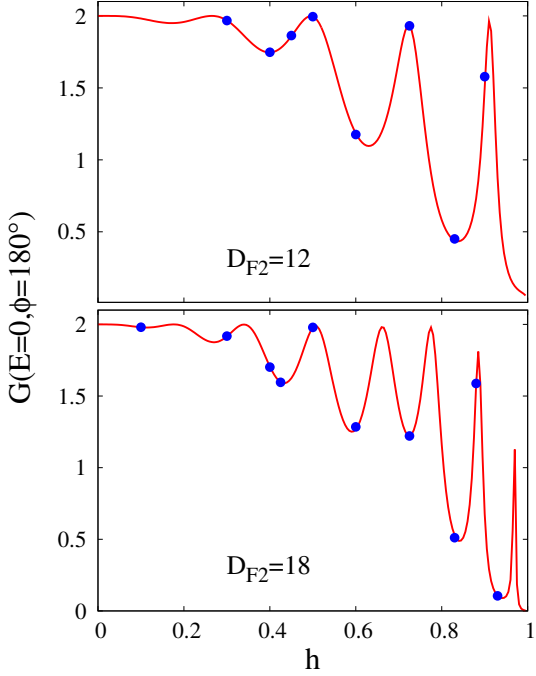} 
\caption{(Color online) Resonance effects in the  forward scattering
conductance at zero bias for  trilayers at $\phi=180^{\circ}$.
In the top panel, the trilayers have same thicknesses as in
the top panels of Fig.~\ref{figure45}, 
and in the bottom panel, they are as in 
the bottom panels of Fig.~\ref{figure45}.
The (blue) dots are the results from our computations and the (red) curves
from Eq.~(\ref{resonance}).}
\label{figure6}
\end{figure}

When the bias is high enough, the tunneling conductance approaches
its normal state value. 
Thus, one can extract
the magnetoresistance from the conductance at  $E=2$.
We only discuss here
the magnetoresistance's qualitative behavior. One can define  
a measure of the magnetoresistance 
as,
\begin{align} 
M_G(E,\phi)\equiv\frac{G(E,\phi=0^{\circ})-G(E,\phi)}{ 
{G(E,\phi=0^{\circ})}}.
\end{align} 
For all results shown in the panels of
Fig.~\ref{figure45}, 
the conductance at $E=2$ decreases with increasing $\phi$,
i.e., 
it is a monotonic function of $\phi$,
the standard behavior for conventional, non-superconducting, 
spin-valves. 
Furthermore,
one can also see that $M_G(E=2,\phi=180^{\circ})$ increases with 
exchange field. Therefore, the behavior of the magnetoresistance 
at large bias is as one would expect in the present 
self-consistent BTK framework.
However, the behavior of $M_G(E=0,\phi=180^\circ)$ that is associated with
the behavior of the ZBC is generally a non-monotonic function of $h$.

We next investigate the high sensitivity of the ZBC to 
$h$ by examining its resonances for two different F widths 
arranged in an AP magnetic configuration ($\phi=180^{\circ}$). 
To do so, we performed an analytic calculation 
of the ZBC in the non-self-consistent framework in situations where
(as discussed in connection with Fig.~\ref{figure3}) 
the results nearly coincide  with those of
self-consistent calculations.  
We find that  the  ZBC 
at $\phi=180^{\circ}$, $G(E=0,\phi=180^\circ)\equiv G_{ZB}$, for a 
given $h$ and $D_{F2}$ 
is: 
\begin{equation}
\label{resonance}
G_{ZB}=\frac{32 k_{\uparrow}^3k_{\downarrow}^3}{A+2\left(h^4-2h^2-2h^2k_{\uparrow}k_{\downarrow}\right)\cos\left[2\left(k_\uparrow-k_\downarrow\right)D_{F2}\right]}.
\end{equation}
The expression for $A$ in Eq.~(\ref{resonance}) is: 
\begin{equation}
a_1\sin^2\left[\left(k_{\uparrow}+k_{\downarrow}\right)D_{F2}\right]
+a_2\left[\cos\left(2k_{\uparrow}D_{F2}\right)-\cos\left(2k_{\downarrow}D_{F2}\right)\right]+a_3,
\end{equation}
where $a_1=4h^2(1-k_{\uparrow}k_{\downarrow})^2$,
$a_2=4h^3$, and $a_3=h^4+(-2+h^2-2k_{\uparrow}k_{\downarrow})^2$.
Here we have omitted the $\pm$ indices for the quasi-particle and quasi-hole
wavevectors, since we are in the zero bias limit. 
In Fig.~\ref{figure6}, we plot Eq.~(\ref{resonance}) as 
a function of $h$ for $D_{F2}=12$ (top panel) and $18$ (bottom panel).
In this zero bias limit, the (blue) circles 
(self-consistent numerical results) are on top of the (red) curves 
(analytic results).
As the thickness of 
the intermediate layer increases, the 
number of resonance maxima and minima increases.
Therefore, the resonance behavior of the ZBC is more sensitive
to $h$ for  larger $D_{F2}$, 
as we have seen in Fig.~\ref{figure45}.
For a given $D_{F2}$, the ZBC 
drops considerably as $\phi$ varies from
$\phi=0^{\circ}$ to $\phi=180^{\circ}$ when $h$ is
near the minimum of the resonance curve (rightmost 
panels of Fig.~\ref{figure45}). 
On the other hand, when $h$ is near the 
resonance maximum (leftmost panels of Fig.~\ref{figure45}), 
the ZBC is a very weak function of $\phi$ 
provided that $h$ is not too strong. 
By examining the denominator of Eq.~(\ref{resonance}), 
we find that the terms involved in $A$ 
are less important than the last term. 
This is because the wavelength $(k_{\uparrow}-k_{\downarrow})^{-1}$
associated with that term is the dominant 
characteristic wavelength in the theory of proximity effects
in F-S structures.\cite{demler,Halterman2002} 
In both panels of Fig.~\ref{figure6}, we see that the ZBC for
$\phi=180^{\circ}$ vanishes in the half-metallic limit. 
To show this analytically, one can  use the conservation of 
probability currents and write down the  relation, valid
when the bias is smaller than the superconducting gap: 
\begin{equation}
\label{probabilitylaw} 
\frac{k^-_{\uparrow 1}}{k^+_{\sigma 1}}|a_{\uparrow }|^2
+\frac{k^-_{\downarrow 1}}{k^+_{\sigma 1}}|a_{\downarrow }|^2
+\frac{k^+_{\uparrow 1}}{k^+_{\sigma 1}}|b_{\uparrow }|^2
+\frac{k^+_{\downarrow 1}}{k^+_{\sigma 1}}|b_{\downarrow }|^2=1. 
\end{equation}
By combining Eq.~(\ref{probabilitylaw}) with Eq.~(\ref{conductance}),
it becomes clear that the subgap conductances arise largely 
from Andreev reflection. 
In the half-metallic limit, 
conventional Andreev reflection is forbidden due to the absence of an
opposite-spin band: this leads to zero ZBC at $\phi=180^{\circ}$. 
Same-spin Andreev reflection 
(see discussion in the paragraph above Eq.~(\ref{f1waveup})) 
is not
allowed in collinear magnetic configurations.
Equation ~(\ref{probabilitylaw}) also reflects 
another important feature of the ZBC: the contributions to $G$ at 
zero bias 
from the spin-up and down channels are identical
except for the weight factor $P_{\sigma}$:
one can prove analytically that 
the sum of first two terms (related to Andreev reflection)
in Eq.~(\ref{probabilitylaw}) is spin-independent. As a result,
the sum of last two terms, related to  ordinary reflection,
is also spin-independent,
and so is the ZBC. 

\begin{figure} 
\includegraphics[width=0.45\textwidth] {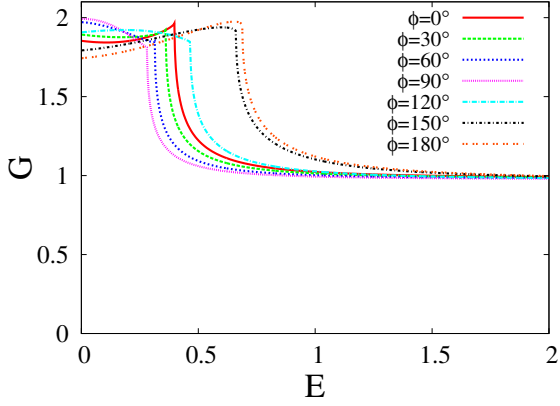}
\caption{(Color online) Forward scattering
conductance of a ${\rm F_1F_2S}$ trilayer with differing magnetic materials
corresponding to exchange fields of
$h_1=0.6$  and $h_2=0.1$. Various magnetic orientations, $\phi$, are considered as shown. Geometry and
other parameters are as in the top panels of Fig.~\ref{figure45}.} 
\label{figure11a}
\end{figure}

\begin{figure*}
\includegraphics[width=0.9\textwidth] {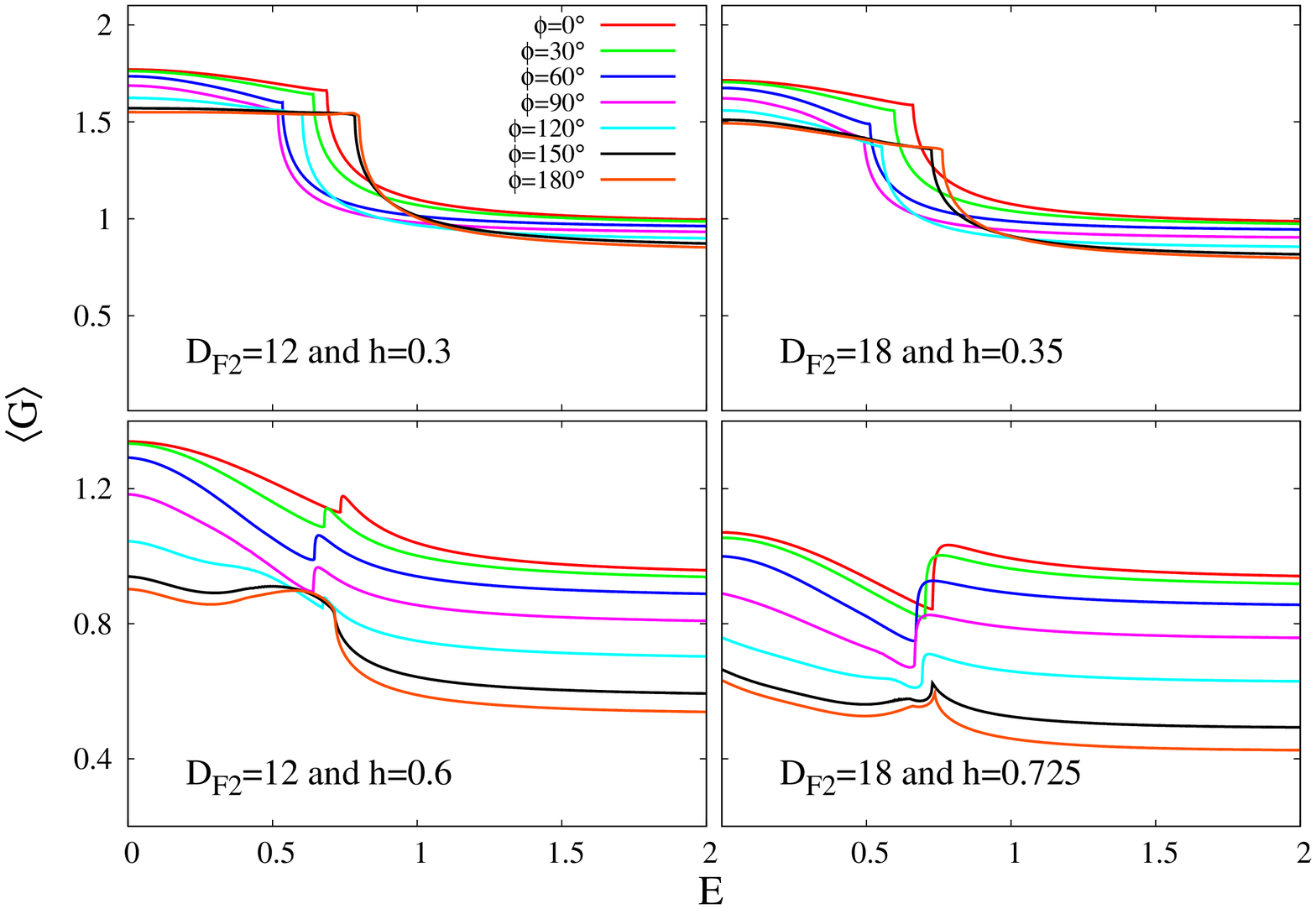} 
\caption{(Color online) Bias
dependence of the angularly averaged conductance 
of ${\rm F_1F_2S}$ trilayers for several angles $\phi$ (see legend).
In the left panels, $D_{F1}=10$,and $D_{S}=180=1.5 \Xi_0$,
as in the top panels of Fig.~\ref{figure45}.
In the right panels, $D_{F1}=10$ as in the bottom
panels of Fig.~\ref{figure45}. 
In all cases, 
 the ZBC decreases with increasing $\phi$. 
}
\label{figure7}
\end{figure*}

We briefly consider here one example 
where the two F materials in the trilayers have
different field strengths.
In this example all the thicknesses and the coherence length are 
as in the top panels of Fig.~\ref{figure45}. 
In Fig.~\ref{figure11a}, we plot
the forward scattering conductance for several $\phi$ at $h_1=0.6$ and
$h_2=0.1$.
One can quickly identify that the ZBC here is a non-monotonic function
of $\phi$ with it maximum at the orthogonal relative 
magnetization angle, $\phi=90^{\circ}$. 
In contrast, results  at equal exchange field strengths
usually demonstrate monotonic 
behavior, as previously shown. 
However,
many features are still the same, such as the formation of a
FBCP when $\phi>90^{\circ}$. For $\phi=0^{\circ}$ and $\phi=30^{\circ}$,
the conductance curves are not monotonically decreasing, as was the 
case at $h_1=h_2$. There, when $h_1=h_2$ and $\phi<90^{\circ}$,
we always see monotonically decreasing behavior 
because the
scattering effect due to misoriented magnetizations is 
not as great as at $\phi>90^{\circ}$.
Also, when $h_1 \neq h_2$, we have to include 
in our considerations another
scattering effect that comes from the mismatch between
$k_{\uparrow1,(\downarrow 1)}^{\pm}$
and $k_{\uparrow 2,(\downarrow 2)}^{\pm}$.
Specifically, when $\phi=0^{\circ}$, the only important
scattering effect is that due to  mismatch from $h_1\neq h_2$ and 
it leads to 
suppression of the ZBC at $\phi=0^{\circ}$.
However, we see that the scattering due to the misoriented magnetic 
configuration ($\phi\neq 0^{\circ}$) compensates the effect of
mismatch from $h_1 \neq h_2$ and ZBC is maximized when $\phi=90^{\circ}$.
Qualitatively, one can examine Eqs.~(\ref{f2spinor1}) and (\ref{f2spinor2}) 
and verify that the spinor at $\phi=90^{\circ}$ is composed of
both pure spin-up and spin-down spinors with equal weight, apart
from phase factors. As a result, the scattering effect due to mismatch from
$k_{\uparrow1,(\downarrow 1)}^{\pm}$
and $k_{\uparrow 2,(\downarrow 2)}^{\pm}$ is reduced.
We also verified that, when the strength of $h_2$ is increased towards 
$h_1$, the locations for the maximum of the ZBC($\phi$) curves
gradually move from $\phi=90^{\circ}$ at $h_2=0.1$ to $\phi=0^{\circ}$
at $h_2=0.6$.

\subsubsection{Angularly averaged conductance}

We now present results for the angularly averaged conductance, 
 $\langle G \rangle$ as defined in Eq.~(\ref{aaG}). The details of the
angular averaging are  explained under Eq.~(\ref{snell}).
The angularly averaged 
conductance
is relevant to a much wider range of experimental results
than the forward conductance, which is
relevant strictly only for some point
contact experiments. This
is particularly true if one recalls  
that the critical angle $\theta_{c\sigma}$
and the weight factor for angular averaging in Eq.~(\ref{aaG})
used in this work can be modified based on a real experimental set-up or
on the geometry of the junction. 

In Fig.~\ref{figure7}, we present  results for
$\langle G \rangle$ at $D_{F2}=12$ (left panels) and
$D_{F2}=18$ (right panels). All curves
are 
obtained with $D_{F1}=10$ and $D_S=180=1.5 \Xi_0$ 
at the values of $h$ indicated
in each panel. Results are plotted over the entire range of $\phi$ values. 
The CB values obtained for  $\langle G \rangle$ are again 
non-monotonic functions of $\phi$ and the non-monotonicity matches that of 
the saturated pair amplitudes, for the  reasons previously given.
The CB values for $\langle G \rangle$
in these cases are the same as those for the forward scattering conductance. 
One can also see that the resonance phenomenon is washed out
in the angularly averaged conductance. For example, the resonance curve in 
the top panel of Fig.~\ref{figure6} tells us 
that $h\approx 0.3$ and $h\approx 0.6$ correspond respectively 
to a resonance maximum and minimum of the ZBC in the
forward scattering $G$.
However, in the top left panel of Fig.~\ref{figure7}, the
ZBC is no longer a weak function of $\phi$ and it  gradually decreases
when $\phi$ is increased.
Near the resonance minimum, $h=0.6$,  bottom left panel 
of Fig.~\ref{figure7},
we can see a trace of the appearance of the FBCP when $\phi$ is above 
$90^{\circ}$. This FBCP in $\langle G \rangle$ is not as prominent as 
that in the forward scattering $G$, due to the averaging. 

The magnetoresistance measure $M_G(E=2,\phi)$ is 
larger for $\langle G \rangle$ 
than for the forward scattering conductance. 
For example, $M_G (E=2,\phi=180^{\circ})$ in the
forward scattering conductance for $h=0.6$ and $D_{F2}=12$ is half of that
in $\langle G \rangle$. As for the zero bias magnetoresistance
$M_G(E=0,\phi=180^{\circ})$ in $\langle G \rangle$, it is of about 
the same order as $M_G(E=2,\phi=180^{\circ})$ and it does not depend on
where it is located in the resonance curve, Fig.~\ref{figure6} 
(recall that $M_G(E=0,\phi=180^{\circ})$
for the forward scattering conductance almost vanishes at the  
resonance maximum).  

In the right panels of Fig.~\ref{figure7}, we plot results
for a larger $D_{F2}$ with values of 
$h=0.35$ (near a resonance maximum) and $h=0.725$ (near a resonance minimum). 
They share very similar features with the thinner $D_{F2}$ case 
in the left panels. However, for $h=0.725$, we see that the ZBC values at
different $\phi$ shrink to almost or less than unity and they are just 
barely higher than the conductance at $E=2$ because 
the contributions from Andreev reflection are strongly suppressed in
such a high exchange field.
\begin{figure*}
\includegraphics[width=0.9\textwidth] {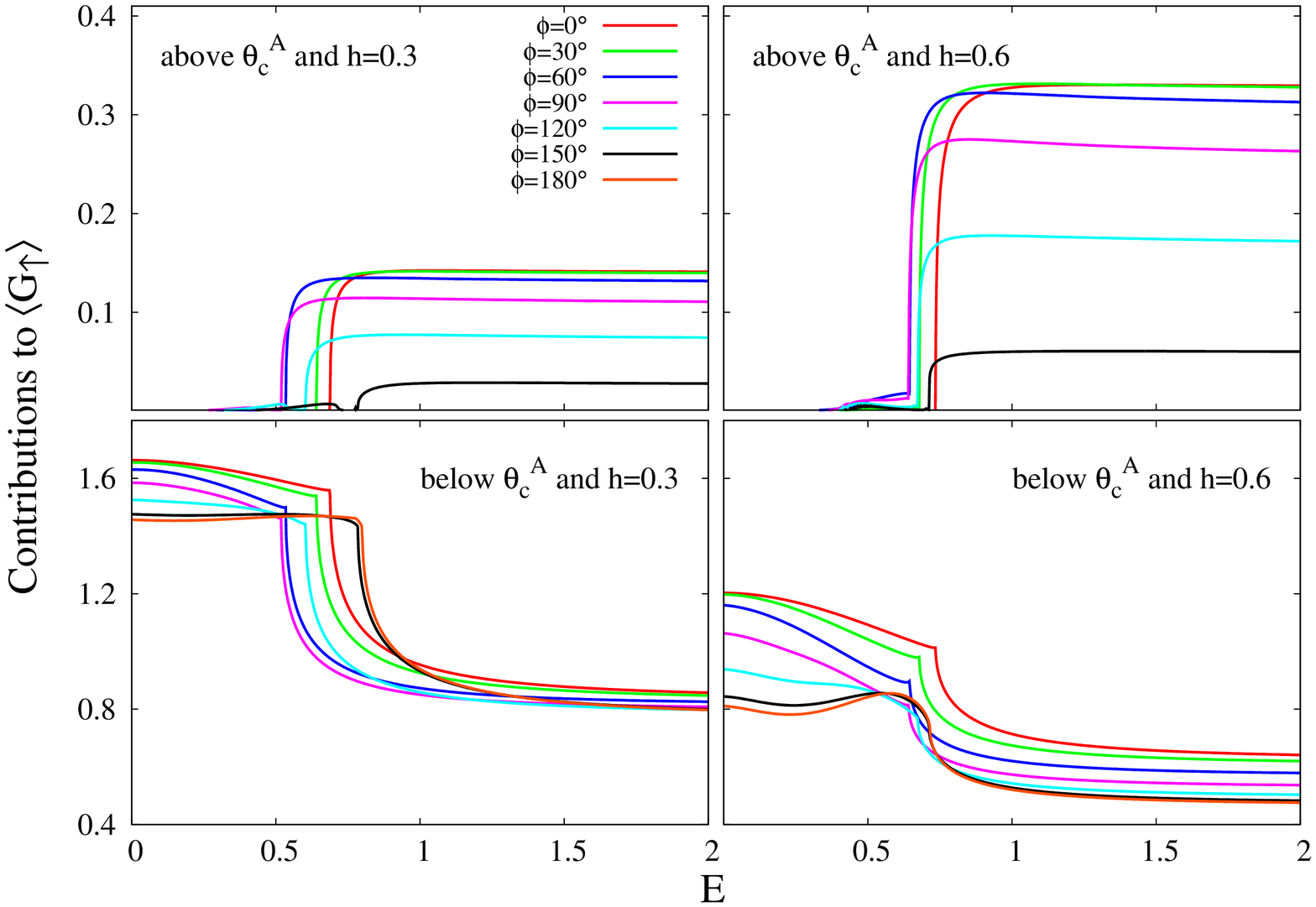} 
\caption{(Color online) Contributions (see text) to the
spin-up angularly averaged 
conductance, $\langle G_{\uparrow}\rangle$,  from angular
ranges above (top panels) and below (bottom panels) the
Andreev critical angle $\theta_c^A$.
Several values of $\phi$ are considered, as indicated. The top panel
results at $\phi=180^\circ$ are vanishingly small.
The geometric and exchange field values are 
as in the left panels of Fig.~\ref{figure7}. 
For the top panels, the plotted values at $E=2$ decrease with 
increasing $\phi$. For the bottom panels, their values  
at both $E=0$ and $E=2$ decrease with increasing $\phi$. 
}
\label{figure8}
\end{figure*}
Another important feature in the angularly averaged results for higher
exchange fields (bottom panels in Fig.~\ref{figure7}) 
is the existence of cusps at 
the CB. 
To understand the formation of these
cusps, we analyze $\langle G \rangle$ by dividing the contribution
from all angles into two ranges: the range above and the range below 
the conventional Andreev critical angles $\theta_c^A$ [see discussion
below Eq.~(\ref{andreevr})].
Consider e.g., the case of spin-up incident
quasi-particles. When
$\theta_c^A\equiv\sin^{-1}[\sqrt{(1-h)/(1+h)}]<\theta_i<
\sin^{-1}[\sqrt{{1}/(1+h)}]$,
the conventional Andreev reflected waves become evanescent while 
the transmitted waves are still traveling waves above the CB. When 
$\theta_i>\sin^{-1}[\sqrt{{1}/(1+h)}]$, both the 
conventional Andreev
reflected waves and the transmitted waves become evanescent. 
Here, $\theta_{c\uparrow}=\sin^{-1}[\sqrt{{1}/(1+h)}]$ 
is the upper limit in Eq.~(\ref{aaG}).

The case of spin-down
incident quasi-particles is trivial, because the 
dimensionless incident momentum 
is $\sqrt{1-h}$ which is less than both the conventional Andreev reflected
wavevector, $\sqrt{1+h}$, and the transmitted wavevector, 
(unity in our 
conventions). Therefore, all the reflected and transmitted waves above the CB 
are traveling waves. As a result, we should consider all possible 
incident angles
and the upper limit of Eq.~(\ref{aaG}) is $\pi/2$.
Let us therefore focus on the nontrivial
spin-up component of $\langle G \rangle$. In Fig.~\ref{figure8}
we separately  plot the contributions
to $\langle G_{\uparrow}\rangle$ from 
angles in the range above $\theta_c^A$ (top panels) and 
below (bottom panels) for the field values and
geometry in the left panels of Fig.~\ref{figure7}, in particular $D_{F2}=12$.
These contributions we will denote as
$\langle G_{\uparrow}(E)\rangle_{above}$ and
$\langle G_{\uparrow}(E)\rangle_{below}$ respectively.
The  
$\langle G_{\uparrow}(E)\rangle_{below}$ contributions, 
in the bottom
panels of Fig.~\ref{figure8} are, for both $h=0.3$ and $h=0.6$,  similar
to the  result for their total forward scattering counterpart  
(see Fig.~\ref{figure3} and
the top right panel of Fig.~\ref{figure45}). 
Of course, the angular averaging leads to a smearing of  the
pronounced features  originally in the forward scattering $G$. 
Qualitatively, the similarity comes from the propagating
nature of all possible waves  except the transmitted
waves below the CB when $\theta_i<\theta_c^A$.
Therefore, the forward scattering $G$ is just 
a special example with the incident angle perpendicular to
the interface.

In the subgap region, the contribution to
$\langle G_{\uparrow}(E)\rangle_{above}$
is vanishingly small
although  small humps appear when the exchange fields in the two F layers
are non-collinear, i.e., $\phi\neq0,\pi$. These small humps are 
generated by
the process of anomalous, equal-spin Andreev reflection. This 
process is possible 
in trilayers because, in a non-collinear magnetic configuration,
a spin up quasiparticle 
can Andreev reflect as a spin-up hole. This can be seen from the matrix 
form of the BdG
equations, Eq.~(\ref{bogo}).
The occurrence of anomalous Andreev reflection leads
to some important physics which we shall discuss
in the next sub-subsection. One can see from Fig.~\ref{figure8}, that
when the exchange fields are strictly parallel 
or anti-parallel to each other, anomalous Andreev reflection does not
arise.

Above $\theta_c^A$, the conventional Andreev-reflected wave is evanescent
and it does not contribute to $\langle G_\uparrow \rangle$.
When the bias is above the saturated pair amplitude, contributions to 
$\langle G_\uparrow \rangle$ from the upper range
are provided by both the transmitted
waves and by anomalous Andreev reflected waves.
Recall that ordinary  transmitted waves are propagating when $E$ is greater 
than the saturated pair amplitudes. 
We also see that $\langle G_\uparrow \rangle_{above}$ decreases 
with increasing $\phi$. At $\phi=180^{\circ}$, $\langle G_\uparrow \rangle$
is vanishingly small due to the effect of a large mismatch from
the anti-parallel exchange field.
Note also that the contribution from above $\theta_c^A$ is less in the 
$h=0.3$ case than at $h=0.6$. This is mainly due to a smaller 
fraction of states at $h=0.3$ 
with incident angles larger than $\theta_c^A$. 
On the other hand, the contribution from below 
$\theta_c^A$ is larger in the $h=0.3$ case.
The increase of $\langle G_\uparrow \rangle_{above}$ and the decrease 
of $\langle G_\uparrow \rangle_{below}$ from $h=0.3$ to $h=0.6$ gives rise
to the cusp at the CB, when  adding these two contributions together.

\subsubsection{Anomalous Andreev reflection}
As we have seen, 
equal-spin (anomalous) Andreev reflection (ESAR) can be generated 
when the 
magnetic configuration is non-collinear. 
We have previously shown that  conventional Andreev 
reflection is forbidden when $\theta_i>\theta_c^A=
\sin^{-1}(\sqrt{(1-h_1)/(1+h_1)})$. Thus,  $\theta_c^A$ 
vanishes in the half-metallic limit. In that case,
conventional Andreev reflection is not allowed for any incident angle 
$\theta_i$ and the subgap $\langle G_\uparrow \rangle$ arises only from 
ESAR.  
For this reason, in this sub-subsection we present 
results for a trilayer structure that consists of one half-metal ($h_1=1$) and
a much weaker ($h_2=0.1$) ferromagnet. The weaker ferromagnet 
serves the purpose of generating ESAR. 
A somewhat similar example that has been
extensively discussed in the literature is 
that of half metal-superconductor bilayers
with spin-flip interface.\cite{eschrig2003,visani,linder2010,niu} 
There the spin-flip interface plays the same
role as the weaker ferromagnet here. Another interesting phenomenon
also related to ESAR is the induction of triplet  
pairing correlations in F-S structures.\cite{Halterman2007,hv2p,ji,cko} 
To induce this
type of triplet pairing, F-S systems must be in a 
non-collinear magnetic configuration such as 
${\rm F_1F_2S}$ or ${\rm F_1SF_2}$ trilayers with $\phi\neq 0,\pi$.
Hence, the mechanism behind induced triplet pairing correlations is
also responsible for ESAR and these two phenomena are closely related.

\begin{figure} 
\includegraphics[width=0.45\textwidth] {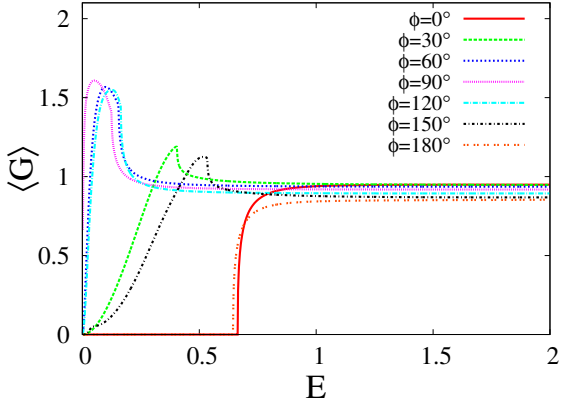}
\caption{(Color online) The angularly averaged conductance
of ${\rm F_1F_2S}$  trilayers with exchange field $h_1=1$ 
and $h_2=0.1$  for several values of $\phi$. See text for discussion.}
\label{figure9}
\end{figure}

\begin{figure}  
\includegraphics[width=0.45\textwidth] {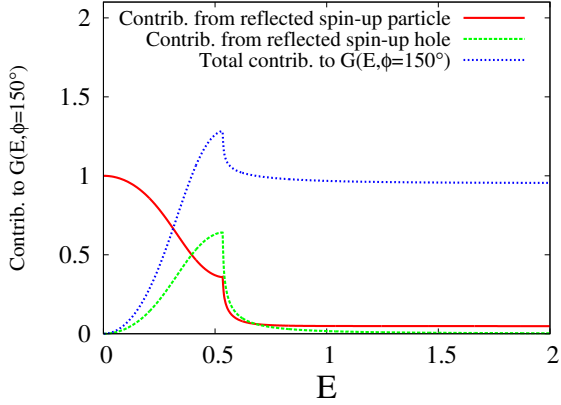}
\caption{(Color online) Contributions to $G(E,\phi=150^\circ)$, 
computed for the parameter values used in Fig.~\ref{figure9},
from the spin-up quasiparticle and spin-up quasihole ESAR
(see text for discussion).  
The total $G$ is also shown. }
\label{figure10}
\end{figure}
In Fig.~\ref{figure9}, we plot the $\langle G \rangle$ of this particular 
system for several $\phi$. The geometrical
parameters are again $D_{F1}=10$, $D_{F2}=12$, and $D_{S}=180$. 
We have 
$\langle G \rangle=\langle G_{\uparrow} \rangle$
because the weight factor $P_{\downarrow}=0$ in this half metallic case.
For $\phi=0^{\circ}$ and  
$\phi=180^{\circ}$  the CB value is about 0.65 and,
below the CB (in the subgap region), 
$\langle G \rangle$ vanishes because the conventional Andreev 
reflection is completely suppressed and ESAR is not allowed
in the collinear cases.  
For $\phi=30^{\circ}$ and $\phi=150^{\circ}$, the CB is near 0.4
and 0.5 respectively and all of the subgap $\langle G \rangle$ 
is due to ESAR. The CB values for $\phi=60^{\circ}$, $90^{\circ}$,
and $120^{\circ}$ are 0.15, 0.12, and 0.15. For these three
angles, a FBCP clearly forms, arising from the ESAR 
in the subgap region. 

To examine 
the conductance in the subgap
region, which is in this case 
due  only to ESAR, we choose the $\phi=150^{\circ}$ angle and
plot, in Fig.~\ref{figure10},  
the contributions to $G$  
(for this case $G$  and $\langle G \rangle$ are very similar) 
from the 
reflected spin-up particle and 
the reflected spin-up hole wavefunctions. 
The spin-down particle and spin-down hole wavefunctions
are evanescent and do not contribute to the conductance.
Thus, Eq.~(\ref{conductance}) reads
$G=
1+{(k^-_{\uparrow 1}}/{k^+_{\uparrow 1})}|a_{\uparrow }|^2
-
|b_{\uparrow }|^2$. 
The quantities plotted are the second ((green)  curve) and
third ((red) curve, highest at the origin) terms in this expression.
The value of $G$ is also plotted. 
One sees that the reflected ESAR amplitudes decay 
very quickly  above the CB. However,
these  reflected  amplitudes are quite
appreciable in the subgap region. 
In other words, the supercurrent in the subgap region contains 
signatures of the  triplet correlations.
This confirms  the 
simple picture\cite{btk} that above the CB  
the current flowing throughout the junction 
is governed by the transport of 
quasiparticles.  However, below the CB it is dominated by ESAR.


\begin{figure}
\includegraphics[width=0.45\textwidth] {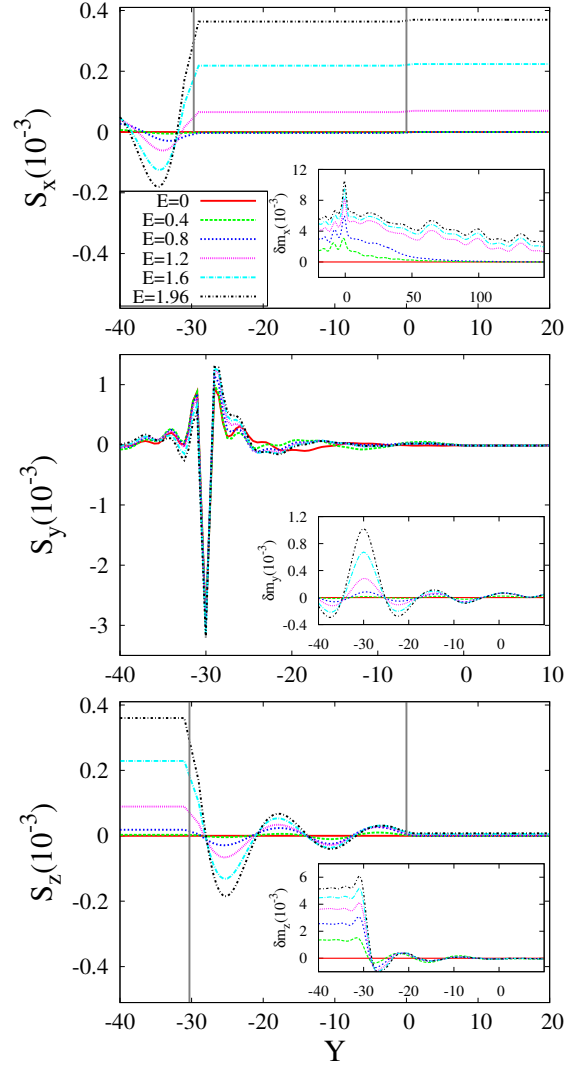} 
\caption{(Color online) The components of the spin current density, 
$S_x$, $S_y$, and $S_z$, calculated from Eq.~(\ref{spincur})  are
plotted vs. $Y\equiv k_Fy$ for several values of the bias $E \equiv eV$
(main panels).
We have $\phi=90^\circ$, $h=0.1$, $D_{F1}=250$, $D_{F2}=30$, $D_S=250=5\Xi_0$.
The F$_2$-S interface is at $Y=0$ and the F$_1$-F$_2$ interface  at
$Y=-30$. 
Vertical lines at these interfaces 
in the top and bottom panels  help locate 
the different regions. 
Only the central portion of the $Y$ range
is included (see text). The  ranges 
included depend on the component.
The insets  show the change 
in each component of the
local magnetization, $\delta \mathbf{m}(E)\equiv \mathbf{m}(E)-\mathbf{m}(0)$,
also as a function of $Y$.  
The values of $E$ are as in the main plot, the ranges included
may be different.
}
\label{figure11}
\end{figure}

\subsubsection{Spin current densities and spin-transfer torques}

Finally, we now report on  
spin-dependent transport quantities, including 
the spin current, the 
spin-transfer torques, and their connections to the local magnetization 
at finite bias. 
An objective here is to 
demonstrate that the conservation law Eq.~(\ref{spinconserve})
which in the steady state is simply:
\begin{equation}
\label{spinsteady}
\frac{\partial}{\partial y} S_i= \tau_i,
\enspace\enspace i=x,y,z, 
\end{equation}
is satisfied in our self consistent calculations for 
F$_1$F$_2$S trilayers. 
We consider  
these spin
dependent quantities in a trilayer with $h=0.1$ and a non-collinear orthogonal magnetic 
configuration, $\phi=90^\circ$. Thus, the internal
field in the outer electrode F$_1$ is along the $z$ axis, while
that in F$_2$ is along $x$. The 
thicknesses are $D_{F1}=250$, $D_{F2}=30$, and $D_S=250=5\Xi_0$.

A set of results is shown 
in Fig.~\ref{figure11}. There, in the three main
plots, we  display the three components of the spin current density, 
computed from Eqs.~(\ref{spincur}) and normalized
as explained below that equation. They are plotted   
as functions of the dimensionless position $Y\equiv k_F y$ for 
several values of the bias $V$, $E \equiv eV$.
The F$_2$-S interface and the F$_1$-F$_2$ interface are located at $Y=0$
and $Y=-30$, respectively. For clarity,
only the range of $Y$ corresponding to the 
''central'' region near the interfaces is included in these plots: the shape
of the curves deeper into
S or F$_1$  can be easily inferred by extrapolation.
From these main panels,  one sees that the current 
is spin-polarized in the $x$-direction (the direction of the exchange 
field in F$_2$)
to the right of F$_1$-F$_2$ interface, including the S region.
Furthermore, $S_x$ is found to be a constant except in the F$_1$ region,
where it exhibits oscillatory behavior. This 
indicates the existence of a non-vanishing,  oscillating spin-transfer torque 
in the F$_1$ layer, as we will verify below. 
We also see that $S_x$ vanishes when the bias is less than the superconducting
gap in bulk S ($E<1$ in our notation). In fact, 
the behavior of $S_x$ with $V$ is similar to that of the ordinary
charge current in an 
N-S tunneling junction with a very strong barrier
where there is no current until $V>\Delta_0$.
This phenomenon is very different from what occurs
in ordinary  spin valves (F$_1$-F$_2$),
where the spin current is not blocked below any finite characteristic bias.

The $S_y$ component, along the normal
to the layers, is shown in the middle main panel of Fig.~\ref{figure11}. It
depends extremely weakly on the bias $E$. It is very
small except near the interface between the  two
ferromagnets but there it is about an order of magnitude larger 
than the other two components. Hence only a somewhat smaller $Y$ range
is shown.
Unlike the $S_x$ and $S_z$ components, 
$S_y$ does not vanish even when there is no bias applied to the trilayer
(the (red) curve in this panel). 
From these observations, one can infer that $S_y$ is largely derived from 
its static part with only a very small contribution from the effect of finite
bias. The emergence of a static spin current is due to the 
leakage of the local magnetization $m_z$ into the F$_2$ layer and of $m_x$ 
into the F$_1$ layer. This explains why the static spin current $S_y$ 
is mostly localized near the F$_1$-F$_2$ interface.
The $S_z$ component (lower panel) is  constant  in the F$_1$ region, 
as one would
expect. 
It oscillates in the F$_2$ region, and vanishes in the S layer. As opposed
to the
$S_x$ component, 
$S_z$ is non-vanishing, although very small, when $E<1$ 
It increases rapidly with bias 
when $E>1$. The oscillatory behavior of $S_z$, again, is related to 
the local spin-transfer torque as we will verify below.

\begin{figure}
\includegraphics[width=0.45\textwidth] {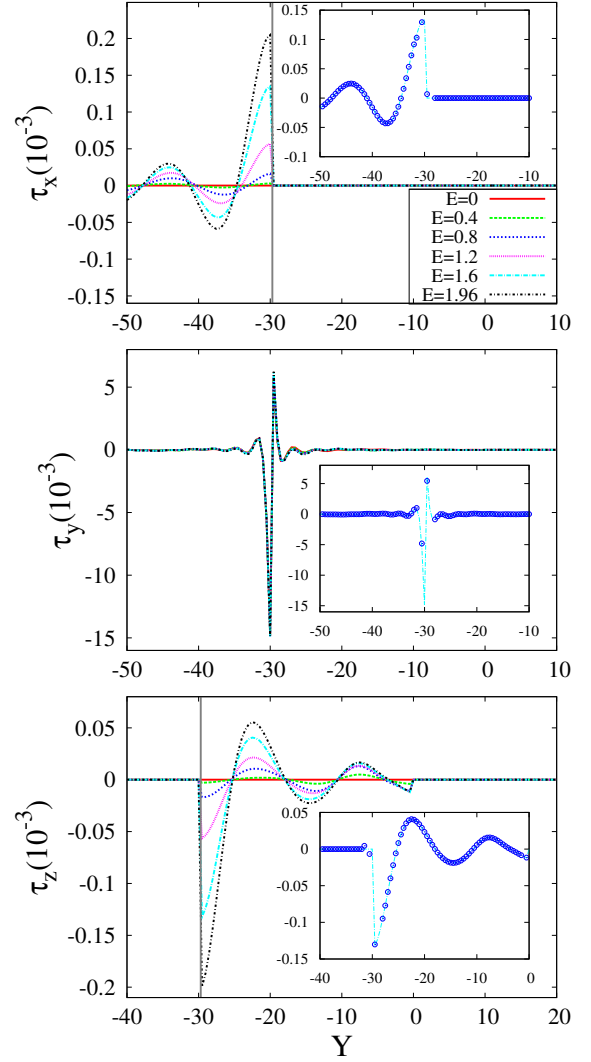}
\caption{(Color online) The components of the spin-transfer torque
$\bm{\tau}\equiv 2\mathbf{m}\times\mathbf{h}$
plotted vs. $Y$ for several  bias values.  All parameters
and geometry are as in
Fig.~\ref{figure11}.
Vertical lines, denoting  interfaces, 
are  in the top and bottom panels. 
The insets show (for bias $E=1.6$) the torque 
((blue) dashed line) 
and the derivative
of the component of spin current density ((blue) circles).
The lines and circles agree, proving that Eq.~(\ref{spinsteady}) holds.}
\label{figure12}
\end{figure} 

We can summarize the behavior of the spin
current  vector, in this $\phi=90^\circ$ 
configuration, as follows:
when $E>1$,
the spin current, which is initially (at the left
side) spin-polarized in the $+z$ direction,
is twisted to the $x$ direction under the action
of the spin torques discussed below,
as it passes through the second magnet, which
therefore acts as a spin filter. 
The current remains then with its spin polarization in
the $+x$ direction as it flows through the superconductor. Thus in this
range of $E$ the trilayer switches the polarization
of the spin current. On the other hand, 
when $E<1$, the small $z$-direction
spin-polarized  current tunneling into the 
superconductor is gradually converted into supercurrent and becomes
spin-unpolarized.

In the insets of the  three panels of Fig.~\ref{figure11}, 
we illustrate the behavior
with bias of the corresponding component of the local magnetization
as it is carried into S.
Specifically, we plot the components of the vector difference between the 
local magnetization with and without bias,
$\delta\mathbf{m}(E)\equiv \mathbf{m}(E)-\mathbf{m}(E=0)$, as a function
of $Y$. The range of $Y$  is chosen to display
the salient aspects of the behavior of this quantity, 
and it is not the same as in the main plots,
nor is it the same for each component. The bias values are the same
as in the main plots, however.
The magnetizations are computed from Eqs.~(\ref{mag}) and normalized
in the usual way,
as discussed below Eqs. (\ref{spincur}). In these units, and at $h=0.1$
the value of the dominant component of ${\bf m}$ in the magnetic
layers is about 0.15.  This scale should be kept in mind.

The behavior of the $x$ component is nontrivial
in the F$_2$ and S regions, and the
corresponding $Y$ range is included in the top
panel inset. When the applied bias is  below the bulk S gap value,
$\delta m_x(E)$ penetrates into the S layer with a decay length 
$\sim\Xi_0=50$.
This decay length is much longer than that found for 
the static magnetization, $\mathbf{m}(0).$\cite{cko} 
When the bias is above the gap,
$\delta m_x(E)$ penetrates  even more deeply into  
the S layer, with a clearly very 
different behavior than for $E<1$. This long-range propagation
is of course consistent with the 
behavior of $S_x$, as $S_x$, the spin
current polarized in the $x$ direction, appears only when $E>1$.
The magnitude of $\delta m_y$ is much smaller than that of $\delta m_x$ or
$\delta m_z$. It peaks near the F$_1$-F$_2$ interface
and that range of $Y$ is emphasized in
the middle inset. Its overall scale monotonically increases with increasing
bias. It damps away from the interface in
an oscillatory manner. 
As to $\delta m_z$, which can conveniently be
plotted in the same $Y$ range, it decays with a very short 
decay length and oscillates 
in F$_2$. The overall damped oscillatory behavior of 
$\delta m_y$ and $\delta m_z$ 
in the F$_2$ region reflects the precession,  
as a function of position, of the spin density
around the local exchange field that points toward the $+x$ direction. 
This phenomenon is well known in  spin-valves.\cite{ralph}
The oscillation periods for $\delta m_y$, $\delta m_z$, $S_x$, and $S_z$ are 
very similar and of the order of $1/(hk_{FS})$. 

Next, we investigate 
the spin-transfer
torque, $\bm{\tau}\equiv 2\mathbf{m}\times\mathbf{h}$. This
quantity, computed from the normalized values of $h$ and ${\bf m}$,
is plotted 
as a function of position in Fig.~\ref{figure12}
for the same system as in Fig.~\ref{figure11}. 
Results are shown   for each of its  three components
in the main panels of the figure. One sees that
at zero bias, $E=0$, both $\tau_x$ and $\tau_z$ vanish identically. 
In the F$_1$ layer, $\tau_x$
increases in magnitude with increasing $E$. It vanishes in 
F$_2$ and in S. The behavior of $\tau_z$ is, as one would
expect, the converse: it vanishes in F$_1$ and S, and its magnitude
increases in F$_2$. The oscillatory behavior of $\tau_x$  and $\tau_z$
is consistent, as we shall see below, with the results for $S_x$
and $S_y$.  The component normal to the layers, $\tau_y$, is nonvanishing
only near the ${\rm F_1}$-${\rm F_2}$ interface, although its peak there attains 
a rather high value, nearly two
orders of magnitude larger than the peak
value of the other components. It is independent of bias, consistent
with the behavior of $S_y$. 

In the insets, we verify, for each component, that Eq.~(\ref{spinsteady})
is satisfied, that is, that our self-consistent methods strictly preserve
the conservation laws in this nontrivial case. (We have already mentioned
that we have verified that the charge or particle current are independent
of $y$). We
specifically consider the bias value $E=1.6$
as an illustration. 
Consider first the top panel inset. There we plot both the $x$ component of 
the spin-transfer torque, $\tau_x$ (blue dashes), 
taken from the
corresponding main plot,
and the derivative of the spin current,
 $\partial S_x/\partial Y$ (blue circles), obtained by numerically
 differentiating the corresponding result in the top panel of
 Fig.~\ref{figure11}.
Clearly, the curves are in perfect agreement. 
(One can easily check that 
with the normalizations and units chosen  there should be no numerical
factor between the two quantities).
In the second panel, the same procedure is performed for the $y$
component, although in this case, because of the very weak dependence
of both $S_y$ and $\tau_y$ on bias, the value of the latter is
hardly relevant. Nevertheless, despite the evident difficulty
in computing the numerical derivative of the very sharply peaked
$S_y$, the agreement is excellent.
For $\tau_z$, its vanishing in the F$_1$ region 
is in agreement with the constant spin current in that layer.
The conservation law Eq.~(\ref{spinsteady}) is verified in the inset
for this component, again at bias $E=1.6$. Just as for the $x$
component, the dots and the line are on top of each other. Thus  
the conservation law for each
component  is shown to be perfectly obeyed.

The results of this sub-subsection can be summarized
as follows: the finite bias leads to 
spin currents. As opposed to the ordinary charge
currents, these spin currents are generally not conserved locally
because of the presence of the spin-transfer torques which act
as  source terms and  are 
responsible for the change of spin-density. But a self-consistent
calculation {\it must} still contain exactly the correct amount of 
non-conservation, that is, Eq.(\ref{spinsteady}) must be satisfied.  It
is therefore of fundamental importance to verify that it is,
as we have.

\section{Conclusions} 
\label{conclusions} 
In summary, we have investigated important 
transport properties of F$_1$F$_2$S trilayers, including 
tunneling conductances and spin transport. To properly take into account 
the proximity effects that lead to a spatially varying pair 
potential, we have incorporated a transfer matrix method into the 
BTK formalism. This allows us to use self-consistent solutions
of the BdG equations. This technique also 
enables us to compute  spin transport quantities 
including spin transfer torque and spin currents. We have 
shown that in F-S bilayers 
the self-consistent calculations 
lead to conductances at the superconducting gap that increase with 
the Fermi wavevector mismatch whereas non-self-consistent ones 
predict they are insensitive to this parameter. 
In F$_1$F$_2$S trilayers, we have found that the
critical bias CB (where tunneling conductance curves drop) for 
different relative magnetization angles, $\phi$, depends on the 
strength of the superconducting order parameter near the interface. 
The angular 
dependence of the critical bias reflects that of the transition 
temperatures $T_c$, which are usually nonmonotonic functions of $\phi$. 
For forward scattering in these F$_1$F$_2$S trilayers, we found that 
the dependence of the zero bias conductance peak (ZBCP) on 
$\phi$ is related to both the strength of the exchange fields
and the thickness of the $F_2$ layers. This remarkable behavior can 
be explained via  quantum interference effects. At the  
resonance minimum, the ZBCP drops significantly and monotonically 
from $\phi=0^\circ$ to $\phi=180^\circ$. On the other hand,
the $\phi$ dependence of the ZBCP is very weak when it is at  its 
resonance maximum. For asymmetric cases where $h_1\neq h_2$, we 
found that the ZBCP is a nonmonotonic function of $\phi$ with 
its value at $\phi=\pi/2$ being the maximum. We have also investigated the 
angularly averaged tunneling conductances, $\langle G \rangle$,
and found that features of resonance effects are then somewhat washed 
out due to the averaging.
However, by studying $\langle G \rangle$ in the subgap regions, 
we found that anomalous (equal spin) Andreev reflection 
(ESAR) arises when $\phi$ corresponds to noncollinear orientations. 
The emergence 
of ESAR is correlated with the well-known induced triplet 
pairing correlations in proximity coupled F-S structures. 
When the outer magnet is a half metal, the  $\langle G \rangle$  signatures
arise 
chiefly from the process of ESAR. 
We have also 
studied the bias dependence of the spin currents and spin 
transfer torques and their general behavior in F$_1$F$_2$S 
trilayers with $\phi=90^\circ$ (the exchange fields in F$_1$
and F$_2$ point toward the $z$ and $x$ directions, respectively).
The spin current components are in general non-conserved 
quantities. The $S_z$ component, parallel to the local exchange 
field in the F$_1$ layer, does not change in the F$_1$ region but 
shows damped oscillatory behavior in the F$_2$ layer and eventually
vanishes in the S region.
However, $S_x$ is a constant throughout the F$_2$ and S regions
and oscillates in F$_1$ layers.
We found that $S_y$ (the component normal to the
layers) depends very weakly  on the bias, and thus its 
spatial dependence arises largely from a static effect. 
The bias dependence of $S_x$ in the S region
is very similar to that of the tunneling charge current
in normal/superconductor systems with high barriers:
$S_x$ vanishes in the subgap regions and arises right above the gap.
The behavior of $\bf{m}$ is consistent with that of $\bf{S}$.
We found that $m_x$, parallel to the local exchange fields in F$_2$,
spreads out over the S regions when the bias is larger
than the superconducting gap.
We have also investigated the bias dependence of the spin transfer torques, 
and we have carefully verified that the appropriate continuity
equation
for the spin current is strictly obeyed in our self-consistent approach.
Our method can be extended to include the effects of interfacial scattering
and wavevector mismatch. It can also be used for further study
of the intricate phenomena associated with  spin transport in these systems.

\acknowledgments Portions of this work were supported by IARPA grant No.
N66001-12-1-2023. CTW thanks the University of Minnesota for a Dissertation
Fellowship. The authors thank I. Krivorotov (Irvine) for helpful
discussions.

\end{document}